\documentclass[A4,titlepage,11pt]{article}

\pdfoutput=1

\usepackage{amssymb,amsmath,amsfonts}
\usepackage{epsfig}
\usepackage{ccaption}
\usepackage{graphicx}
\usepackage{subfigure}

\usepackage[
      colorlinks=true,
      linkcolor=blue,
      urlcolor=blue,
      filecolor=black,
      citecolor=red,
      pdfstartview=FitV,
      pdftitle={},
        pdfauthor={Roberto Emparan, Ryotaku Suzuki, Kentaro Tanabe},
        pdfsubject={},
        pdfkeywords={},
        pdfpagemode=None,
        bookmarksopen=true
      ]{hyperref}

\def\clock{{\count0=\time
           \divide\count0 60
           \ifnum\count0<10 0\fi\the\count0
           \multiply\count0 -60 \advance\count0 \time
           :\ifnum\count0<10 0\fi \the\count0
         }}
\newcommand{\timestamp}{{\small\vbox{\hbox{\tt\jobname.tex}
\hbox{\the\day/\the\month/\the\year, \clock}}}}

\newcommand{\mc}[1]{\mathcal{#1}}

\newcommand{\beq}{\begin{equation}}
\newcommand{\eeq}{\end{equation}}
\newcommand{\bea}{\begin{eqnarray}}
\newcommand{\eea}{\end{eqnarray}}
\newcommand{\beqa}{\begin{eqnarray}}
\newcommand{\eeqa}{\end{eqnarray}}

\newcommand{\sR}{\mathsf{R}}

\numberwithin{equation}{section}

\begin{document}
\begin{titlepage}
\centerline{\LARGE \bf Static Gauss-Bonnet Black Holes at Large $D$}

\vskip 1.6cm
\centerline{\bf Bin Chen$^{a,b,c}$ and Peng-Cheng Li$^{a}$}
\vskip 0.5cm
\centerline{\sl $^{a}$Department of Physics and State Key Laboratory of Nuclear Physics and Technology,}
\centerline{\sl Peking University, No.5 Yiheyuan Rd, Beijing 100871, P.R. China\footnote{Email: bchen01@pku.edu.cn,\, wlpch@pku.edu.cn}}
\smallskip
\centerline{\sl $^{b}$Collaborative Innovation Center of Quantum Matter, No. 5 Yiheyuan Rd,}
\centerline{\sl  Beijing 100871, P. R. China}
\smallskip
\centerline{\sl $^{c}$Center for High Energy Physics, Peking University, No.5 Yiheyuan Rd,}
\centerline{\sl  Beijing 100871, P. R. China}
\smallskip
\vskip 0.5cm

\vskip 1.2cm
\centerline{\bf Abstract} \vskip 0.2cm
\noindent
We study the static   black holes in the  large $D$ dimensions in the Gauss-Bonnet gravity with a cosmological constant, coupled to the Maxewell theory. After integrating the equation of motion with respect to the  radial direction,  we obtain the effective equations at large $D$ to describe the nonlinear dynamical deformations of the  black holes. From the perturbation analysis on the effective equations, we  get the analytic expressions of the frequencies for the  quasinormal modes  of charge and scalar-type perturbations. We show that for a positive Gauss-Bonnet term, the  black hole could become unstable only if the  cosmological constant is positive,  otherwise the black hole is always stable. However, for a negative Gauss-Bonnet term, we find that the black hole could always be unstable. The instability of the black hole depends not only on the cosmological constant and the charge, but also significantly on the  Gauss-Bonnet term. Moreover, at the onset of instability there is a non-trivial static zero-mode perturbation, which suggests the existence of a new non-spherically symmetric solution branch. We construct  the non-spherical symmetric static  solutions of the large $D$ effective equations explicitly.

\end{titlepage}
\pagestyle{empty}
\small
\normalsize
\newpage
\pagestyle{plain}
\setcounter{page}{1}

\section{Introduction}

Recently it has been found that black hole physics in Einstein's gravity can be efficiently investigated by using the $1/D$ expansion in the near region of the black hole. The large $D$ expansion was first proposed by R. Emparan, R. Suzuki and K. Tanabe (EST) in \cite{EST13} and subsequently developed in a series of papers\cite{EST14}. The essence in the large $D$ expansion  is that when the spacetime dimension is sufficiently large $D\to \infty$, the gravitational field of a black hole is strongly localized near its horizon due to the dominant radial gradient of the gravitational potential. As a result, for the decoupled quasinormal modes \cite{EST1406} the black hole can be effectively taken as a surface or membrane embedded in the  background spacetime\cite{ESTT, ST, BDMMS, BMMT, Emparan:2015gva, Tanabe:2015hda}. The membrane is described by the way it is embedded into the background spacetime, and its  nonlinear dynamics is determined by the effective equations obtained by integrating the Einstein equations in the radial direction.  Then the frequencies of the decoupled  quasinormal modes of the black hole solutions can be obtained by performing the perturbation analysis of the effective equations. Furthermore, by solving the effective equations with different embedding of the membrane, one can construct different black hole solutions such as non-uniform black string \cite{ST:Nonuniform} and black rings \cite{Tanabe:2015hda}, and  study numerically the final fate of their evolutions due to the instability  \cite{Emparan:2015gva, EILST:Hydro, RVE:Onbrane}.

To understand the large $D$ expansion method better, it is valuable to extend the study to other gravity theories.
One of most interesting generalization of  Einstein gravity in higher dimensions is the Lovelock higher-curvature gravity of various orders. The most attractive feature of  the Lovelock gravity is that its equations of motion  are still the second order differential equations such that the fluctuations around the vacuum do not have ghost-like mode. Among all
the Lovelock gravities, the second-order Lovelock gravity, the so called Einstein-Gauss-Bonnet gravity, is of particular interest. It includes the quadratic terms of the curvature tensors which appear as the leading-order correction in the low energy effective action of the heterotic string theory\cite{Zwiebach, Boulware}. The exact spherically symmetric black hole solution of the Gauss-Bonnet gravity theory was  discovered by Boulware and Deser\cite{Boulware} and independently by Wheeler\cite{Wheeler}(but without a cosmological constant). The construction of the Gauss-Bonnet black hole solution was later generalized to the case with an electric charge\cite{Wiltshire, Wiltshire1988}.

The study of the Gauss-Bonnet black holes at large $D$ was initiated in \cite{CFLY}. The quasi-normal modes of the black holes in the large $D$ limit have been computed carefully. It was found that the decoupled quasinormal modes, which characterize the information of the black hole, share the similar features as the ones in the Einstein gravity. The computation of the quasinormal modes in \cite{CFLY} relied on the master equations of the fluctuations.
On the other hand, it turns out to be more efficient  at large $D$ to study the black hole dynamics using the effective theory.  In this paper, we would like to discuss  the large $D$ effective theory of the Gauss-Bonnet black holes and study their instabilities.

The stability of the black hole under the perturbation is an important issue. It has been investigated for the Gauss-Bonnet black holes since their findings.
 For the asymptotically flat Gauss-Bonnet black holes it was found in \cite{Dotti, Gleiser} that such black holes are unstable against gravitational perturbations in  five and six dimensions but become stable in higher dimensions \cite{Konoplya}. For a Gauss-Bonnet black hole with a positive cosmological constant, it was shown in \cite{Cuyubamba} that the black holes  becomes unstable in  $D\geq 5$ dimensions at sufficiently large values of the cosmological constant. In \cite{CFLY}, the instability of the asymptotically flat Gauss-Bonnet black holes at large $D$ has been discussed in two interesting limits. One limit is that the Gauss-Bonnet term  can be treated as a small correction to the Einstein gravity, and the other one is to let the Gauss-Bonnet term be dominant. In both cases the Gauss-Bonnet black holes are found to be  stable. One unsolved question is  the instability of the  black hole  for an arbitrary Gauss-Bonnet term. Moreover, the studies on the stability of the  black hole have been focused on the case with a positive Gauss-Bonnet term, it would be interesting to consider the case with a negative  Gauss-Bonnet term.

 In the presence of the cosmological constant and the charge, the stability of the black hole at large $D$ can be very different.
 In \cite{Tanabe15}, it was shown  that the de Sitter Reissner-Nordstrom black hole in the Einstein gravity  becomes unstable against scalar-type gravitational perturbation when the  charge is sufficient large. Moreover, it was found that there is a non-trivial zero-mode static perturbation at the threshold of the  instability. The existence of such zero-mode indicates that there must be a non-spherical symmetric solution branch of static charged de Sitter black holes, whose specific form can be constructed by solving  the large $D$ effective equations. In this work,  we  would like to investigate the effect of the cosmological constant and/or the charge on the stability of the Gauss-Bonnet black holes. 


The remaining part of the paper is organized  as follows. In Section \ref{secde} we briefly review  the (Anti-)de Sitter charged Gauss-Bonnet black holes. In Section \ref{largeD} we derive the  large $D$ effective equations for the black holes, and in Section \ref{instability} we perform the stability analysis of various black holes in the theory with a positive Gauss-Bonnet term. In Section $\ref{appendixA}$ we extend the stability analysis of the black holes to the case with a negative Gauss-Bonnet term.
We end with some conclusion and discussions in Section \ref{summary}.

\section{(Anti-)de Sitter charged Gauss-Bonnet black holes}
\label{secde}

Let us consider the Gauss-Bonnet gravity with a positive cosmological constant, coupled to the Maxwell theory. Its action is given by
\beq\label{eq2s1}
S=\frac{1}{16\pi G}\int d^dx\sqrt{-g}\biggl(R+\alpha(R_{\mu\nu\lambda\delta}R^{\mu\nu\lambda\delta}-4R_{\mu\nu}R^{\mu\nu}+R^2)-\frac{(D-1)(D-2)}{L^2}-\frac{1}{4}F_{\mu\nu}F^{\mu\nu}\biggl).
\eeq
Here $\alpha$ is the the Gauss-Bonnet coefficient. It is positive definite and inversely proportional to the string tension in the heterotic string theory\cite{Boulware}. However,   we do not restrict ourselves to the case $\alpha\geq0$ and allow it to be free parameter in this paper. The cosmological constant can be negative, which gives the black hole solution in AdS, but the form of the solution is similar.

From the action, we obtain the  equations of motion for the metric
\beqa
R_{\mu\nu}-\frac{1}{2}g_{\mu\nu}R&=&-\frac{(D-1)(D-2)}{2L^2}g_{\mu\nu}+\alpha\biggl(\frac{1}{2}g_{\mu\nu}(R_{\sigma\gamma\lambda\delta}R^{\sigma\gamma\lambda\delta}-4R_{\lambda\delta}R^{\lambda\delta}+R^2)\nonumber\\
&\,&-2RR_{\mu\nu}+4R_{\mu\gamma}R^{\gamma}_{\,\,\nu}-4R^{\gamma\delta}R_{\gamma\mu\nu\delta}-2R_{\mu\gamma\delta\lambda}R_{\nu}^{\,\,\gamma\delta\lambda}\biggl)\nonumber\\
&\,&+\frac{1}{2}\biggl(F_{\mu\rho }F_{\nu}^{\,\,\rho}-\frac{1}{4}g_{\mu\nu}F_{\rho\sigma}F^{\rho\sigma}\biggl),
\eeqa
and the Maxwell equations
\beq
\nabla^\mu F_{\mu\nu}=0,
\eeq
where $F_{\mu\nu}=\partial_\mu A_\nu-\partial_\nu A_\mu$.
The spherically symmetric charged de Sitter Gauss-Bonnet black hole has the metric \cite{Wiltshire1988}
\beq\label{eq2s4}
ds^2=-f(r)dt^2+f^{-1}(r)dr^2+r^2d\Omega^2_{D-2},
\eeq
where
\beq
f(r)=1+\frac{r^2}{2\tilde{\alpha}}\biggl(1-\sqrt{1+\frac{64\pi G\tilde{\alpha}M}{(D-2)\Omega_{D-2}r^{D-1}}-\frac{2\tilde{\alpha}Q_e^2}{(D-2)(D-3)r^{2D-4}}+\frac{4\tilde{\alpha}}{L^2}}\biggl),
\eeq
and the Maxwell field\footnote{Note that here we use the convention in \cite{MP} . It seems that there is a typo for the field strength $F_{\mu\nu}$ in eq.(1.4b) in \cite{Wiltshire1988}. The correct form should be $F=\frac{Q}{4\pi r^{D-2}}dt\wedge dr$.}
\beq
A_\mu dx^{\mu}=\frac{Q_e}{(D-3)r^{D-3}}dt.
\eeq
In the following discussion we will assume that the largest root of $f(r)=0$ which corresponds to the outer horizon always exists. In the metric function $f(r)$, $M$ is the mass of the black hole which can be expressed as
\beq
M=\frac{(D-2)\Omega_{D-2}r_+^{D-3}}{16\pi G}(1+\frac{\tilde{\alpha}}{r_+^2}),
\eeq
with
\beq
\tilde{\alpha}=\alpha(D-3)(D-4).
\eeq
 It is easy to see that when $1/L\to0$ and $Q\to0$, $r_+$ is just the horizon radius of the usual Gauss-Bonnet black hole.

In order to discuss the large $D$ expansion more conveniently, we introduce
\beq
n=D-3,
\eeq
and
\beq
\sR=\big(\frac{r}{r_0}\big)^n,
\eeq
where $r_0$ is a constant parameter and can be set to be unity. In order to see the effect of the gauge field on the solution in the large $D$ limit, we should replace $Q_e$  with another   $\mc O(1)$ variable
\beq
\widetilde{Q}=\frac{Q_e}{r_+^n\sqrt{2n(n+1)}},
\eeq
in terms of which the gauge field can be written as
\beq
A_\mu dx^{\mu}=\sqrt{\frac{2(n+1)}{n}}\widetilde{Q}\big(\frac{r_+}{r}\big)^ndt.
\eeq
In terms of $\sR$, at the leading order of $1/n$ expansion, $f(r)$ becomes
\beq\label{exact}
f(\sR)=1+\frac{r_0^2}{2\tilde{\alpha}}\biggl(1-\sqrt{1+\frac{4\tilde{\alpha}}{\sR}\frac{r_+^n(1+\frac{\tilde{\alpha}}{r_+^2})}{r_0^{n+2}}-\frac{4\tilde{\alpha}}{\sR^2}\frac{\widetilde{Q}^2r_+^{2n}}{r_0^{2n+2}}+\frac{4\tilde{\alpha}}{L^2}}\biggl).
\eeq
From the above expression we can see that if we use $Q_e$ instead of $\widetilde{Q}$, at the leading order of $1/n$ expansion the gauge field term disappears. This fact suggests that the variable  $\widetilde{Q}$ is more appropriate  in the large $D$ expansion.

\section{Large $D$ effective equations}
\label{largeD}
In this section, we  derive the large $D$ effective equations for the theory \eqref{eq2s1} and hence study the stabilities of the solutions. First of all, for the spherically symmetric metric solution \eqref{eq2s4} we can make the metric ansatz in terms of the ingoing Eddington-Finkelstein coordinates as
\beq
ds^2=-Adv^2+2(u_vdv+u_zdz)dr-2C_zdvdz+r^2Gdz^2+r^2H^2d\Omega_n^2,
\eeq
where $z$ is the inhomogeneous coordinate and can be interpreted as one of the coordinates of $d\Omega_{n+1}^2$. The generalization to the case with several inhomogeneous coordinates would  be straightforward. It might also be possible to use this metric to describe Gauss-Bonnet black strings with appropriate embedding of the membrane, where $z$ is the  direction the string extends along. 
 In this article we will not pursue this interesting problem further but leave it for future work, instead we just focus on the black hole solutions.

The gauge filed ansatz is
\beq
A_\mu dx^\mu=A_vdv+A_zdz.
\eeq
As explained in \cite{Tanabe15}, the $A_r  dr$ term is omitted due to the fact that it does not contribute to effective equations at the leading order in the $1/n$ expansion. For example in $F_{tr}=\partial_t A_r-\partial_r A_t$, the first term is of $\mc O(1/n)$ if we assume that $A_r=\mc O(1/n)$, but the second term is of  $\mc O(n)$ since $\partial_r=\mc O(n)$.
The functions in the metric and gauge fields generally depend on $(v,\sR,z)$ except that $G$ and $H$  depend only on $z$ because at the asymptotic infinity $\partial_v$ is a Killing vector. In order to do the $1/n$ expansion properly we need to specify the large $D$ behaviors  of these functions. Their large $n$ scalings are respectively
\beq
A,\, u_v,\, A_v=\mc O(1),\quad u_z,\, C_z,\, A_z=\mc O(1/n), \quad G=1+\mc O(1/n).
\eeq

At the leading order, since $\partial_r=\mc O(n)$, $\partial_v=\mc O(1)$ and $\partial_z=\mc O(1)$  the equations of motion only contain $\sR$-derivative so they can be solved by performing $\sR$ integrations. The leading order solutions can be   written as\footnote{Note that the $\mc O(1/n)$ term in $G$ vanishes identically at the leading order in the $1/n$ expansion.}
\beq
A=A_0^2+\frac{u_v^2}{L^2}+\frac{u_v^2}{2\tilde{\alpha}}\Bigg(1-\sqrt{1+\frac{4\tilde{\alpha}}{L^2}+\frac{4\tilde{\alpha}m}{\sR u_v^2}-\frac{4\tilde{\alpha} q^2}{\sR^2 u_v^2}}\Bigg),
\eeq
\beq
A_v=\frac{\sqrt{2}q}{\sR},\quad A_z=-\frac1n\frac{\sqrt{2}p_z\,q}{m\sR},\quad u_v=\frac{A_0}{\sqrt{\frac{1-H'(z)^2}{H(z)^2}-\frac{1}{L^2}}},\quad G=1+\mc O(n^{-2}),
\eeq
\beq
C_z=\frac1n\frac{p_z\,u_v^2}{m}\Biggl[-(\frac{1}{2\tilde{\alpha}}+\frac{1}{L^2})+\frac{1}{2\tilde{\alpha}}\sqrt{1+\frac{4\tilde{\alpha}}{L^2}+\frac{4\tilde{\alpha}m}{\sR u_v^2}-\frac{4\tilde{\alpha} q^2}{\sR^2 u_v^2}} \Biggl].
\eeq
The integration functions in the solutions are the functions of $(v, z)$.
In fact it is easy to see that in the limit $\tilde{\alpha}\to0$  the above expressions reduce to the ones found in \cite{Tanabe15}. Moreover  the effect of the Gauss-Bonnet term only appears in the functions $A$ and $C_z$, without appearing in other functions. In particular $C_z$ has a simple relation with $A$ as \beq
C_z=\frac1n\frac{p_z}{m}(-A+A_0^2),
\eeq
 which has also been found in the Einstein gravity.

 Comparing with the exact solution (\ref{exact}) we can read the physical meanings of the  integration functions appearing in the leading order solutions: $m(v, z)$ and $q(v, z)$ are the mass density and the charge density respectively, $p_z(v, z)$ is a momentum density along $z$ direction, $A_0$ is a constant which is independent of $v$ and $z$, and $u_z$ is a shift vector on a $r=\textrm{const}.$ hypersurface. At the leading order $u_z$ can be gauged away so it vanishes identically.
From the leading order form of the metric it is  easy to find the outer and inner horizon radii of the dynamical black hole, which correspond to the two roots of $A=0$,
\beq\label{R+}
\sR_{\pm}=\frac{m\pm\sqrt{m^2-4\Big(A_0^2+\tilde{\alpha}(\frac{A_0^2}{u_v}+\frac{u_v}{L^2})^2\Big)q^2}}{2\Big(A_0^2+\tilde{\alpha}(\frac{A_0^2}{u_v}+\frac{u_v}{L^2})^2\Big)}.
\eeq

At the next-to-leading order, we could obtain the equations for $m(v, z)$, $q(v, z)$ and $p_z(v, z)$. They can be taken as the effective equations for the charged de Sitter Gauss-Bonnet black holes. These equations are
\beq\label{eff1}
\partial_v q-\frac{u_vH'(z)}{H(z)}\partial_zq+\frac{A_0^2H'(z)}{H(z)}\frac{p_zq}{m}=0,
\eeq
\beq\label{eff2}
\partial_v m-\frac{u_vH'(z)}{H(z)}\partial_zm+\frac{A_0^2H'(z)}{H(z)}p_z=0,
\eeq
\beqa\label{eff3}
\partial_v p_z&-&\frac{u_v H'(z)}{m\,H(z)}\frac{2A_0^2u_v^2(L^2+4\tilde{\alpha})\sR_+-(2\tilde{\alpha}u_v^2+L^2(u_v^2-2A_0^2\tilde{\alpha}))m}{2u_v^2\tilde{\alpha}+L^2(u_v^2+2A_0^2\tilde{\alpha})}\partial_z p_z\nonumber\\
&+&\biggl[1+\frac{2u_v^3((A_0^2L^2+4A_0^2\tilde{\alpha})\sR_+-(L^2+2\tilde{\alpha})m)p_z}{L^2u_v^2+2A_0^2L^2\tilde{\alpha}+2u_v^2\tilde{\alpha}}\frac{H'(z)}{H(z)m^2} \biggl]\partial_z m\\
&-&\bigg[\frac{2u_v^4\tilde{\alpha}+L^2(u_v^4-2A_0^2u_v^2\tilde{\alpha})+4A_0^2(A_0^2L^2+u_v^2)\tilde{\alpha}H(z)^2}{u_v(2u_v^2\tilde{\alpha}+L^2(u_v^2+2A_0^2\tilde{\alpha}))H(z)^2} \nonumber\\
&\,&+\frac{2A_0^2(L^2+4\tilde{\alpha})(-L^2u_v^3+(A_0^2L^2u_v+u_v^3)H(z)^2)\sR_+}{L^2(2u_v^2\tilde{\alpha}+L^2(u_v^2+2A_0^2\tilde{\alpha}))H(z)^2m}-\frac{A_0^2H'(z)}{H(z)}\frac{p_z}{m}\biggl]p_z=0.\nonumber
\eeqa
Besides, in order to describe the embedding of the membrane in the background spacetime, we need to do ``$(D-1)+1$" decomposition on a $r=\textrm{const}.$ surface. Then the momentum constraint of the decomposition  at the leading order gives
\beq\label{momcon}
\frac{d}{dz}\frac{A_0}{\sqrt{\frac{1-H'(z)^2}{H(z)^2}-\frac{1}{L^2}}}=0.
\eeq
The constraint suggests that $u_v$ can be regarded as a constant.

By assuming $m(v,z)=m(z)$, $q(v,z)=q(z)$ and $p_z(v,z)=p_z(z)$ it is straightforward to find a static solution from these effective equations. From \eqref{eff1} and \eqref{eff2},
we obtain
\beq\label{pq}
p_z(z)=\frac{u_v}{A_0^2}m'(z), \qquad q(z)=K m(z),
\eeq
where $K$ is a constant. Plugging (\ref{pq}) into (\ref{eff3}) and setting $m(z)=e^{P(z)}$, we have the equation for $P(z)$
\beqa\label{Pz}
&\,&P''(z)\nonumber\\
&\,&+\Bigg[\frac{A_0^2H(z)}{L^2u_v^2H'(z)}\frac{2u_v^2(A_0^2L^2+u_v^2)(L^2+\tilde{\alpha})S_++2L^2(A_0^2L^2+u_v^2)\tilde{\alpha}}{2A_0^2u_v^2(L^2+4\tilde{\alpha})S_+-L^2u_v^2+2(A_0^2L^2-u_v^2)\tilde{\alpha}}
-\frac{1}{H(z)H'(z)}\Bigg]P'(z)=0,\nonumber\\
\eeqa
where
\beq\label{S+}
S_+=\frac{1+\sqrt{1-4\Big(A_0^2+\tilde{\alpha}(\frac{A_0^2}{u_v}+\frac{u_v}{L^2})^2\Big)K^2}}{2\Big(A_0^2+\tilde{\alpha}(\frac{A_0^2}{u_v}+\frac{u_v}{L^2})^2\Big)}.
\eeq
The functions $H(z)$ and $A_0$ are given through an embedding of the leading order solution into a background spacetime, and they  should satisfy the condition (\ref{momcon}).  As long as we know  $H(z)$ and $A_0$, we can determine the function $m(z)$.

There are two special cases worth noticing. The first one is the  limit $\tilde{\alpha}=0$. In this case  (\ref{Pz}) is used to describe the de Sitter charged black holes. At  the extremal limit $\sR_+=\sR_-$,
(\ref{Pz}) is not valid any more. This fact  can be seen directly  from the denominator of the first term in the square bracket in (\ref{Pz})
\beq
L^4u_v^4H'(z)\sqrt{1-4A_0^2K^2},
\eeq
which  becomes zero in the extremal limit. In other words, the extremal point is singular for the de Sitter Reissner-Nordstrom black holes, similar to the case for the charged black ring\cite{Chen:2017wpf}. In contrast,
 for the Gauss-Bonnet gravity, this singular behavior does not arise, because (\ref{Pz}) is always regular for the physical solution. The other special case is  the limit $A_0=0$. In this case it is obvious (\ref{pq}) is not valid, neither is  (\ref{Pz}). From the effective equations, in this limit we obtain $m(z)=$ const., $q(z)=$ const. and $p_z(z)=0$. Hence in this special case, no inhomogeneous solution exists.

In this static case $\sR_+$ is  proportional to $m(z)$ and  $\partial_v$ becomes the Killing vector. Consequently  the surface gravity $\kappa$ can be expressed in terms of $S_+$ in (\ref{S+})
\beqa\label{kappa}
\kappa &=&\frac{n}{2}\frac{\sR\,\partial_\sR A}{u_v}\Biggl|_{\sR_+}\nonumber\\
&=&C\frac{n}{2u_v}\frac{1}{1+2\tilde{\alpha}(\frac{A_0^2}{u_v^2}+\frac{1}{L^2})},
\eeqa
with
\beq
C=\frac{S_+-2K^2}{S_+^2}.
\eeq
As $C$ is a constant,  so is the surface gravity.

It was interestingly  noted in \cite{ESTT, ST, EILST:Hydro} that  the stationary large $D$  black holes are the solutions of an $elastic$ theory. For example, for the neutral static black holes the effective membrane embedded in the background must satisfy the equation
\beq\label{extrinsic}
\sqrt{-g_{vv}(1-v^2)}\,\mathcal{K}=2\kappa,
\eeq
where $\mathcal{K}$ is the trace of the extrinsic curvature of the membrane, $v$ represents the Lorentz boost on the membrane, $\kappa$ is just the surface gravity of the black hole  and $g_{vv}$ is the redshift factor on the membrane. Especially for the static black holes in the Minkowski background $\sqrt{-g_{vv}}=1$, the membrane amounts to a spherical soap bubble. For Gauss-Bonnet black holes, the static solutions we obtained above are also the solutions of an elastic theory, but the equation is not as simple as the one (\ref{extrinsic}).
For example, if we treat $\tilde{\alpha}$ as a small quantity and keep its leading order, for the neutral case $K=0$, we have
\beq
\sqrt{-g_{vv}}\,\mathcal{K}=\frac{n A_0^2}{u_v}+\frac{n u_v\tilde{\alpha}}{L^4}.
\eeq
However, from (\ref{kappa}) after setting $K=0$ we find
\beq
2\kappa=\frac{n A_0^2}{u_v}+\Big(\frac{n u_v}{L^4}-\frac{A_0^4}{u_v^3} \Big)\tilde{\alpha}.
\eeq
Obviously, these two equations cannot be equal generally unless $A_0=0$, which would lead to a trivial solution. Nevertheless, we still can write the static solutions in a elastic form, that is
\beq
\sqrt{-g_{vv}}\,\mathcal{K}=\textrm{constant}.
\eeq

\section{Instability of de Sitter charged Gauss-Bonnet black hole}
\label{instability}

By embedding  $H(z)$ and $A_0$ into the de Sitter spacetime, the de Sitter charged Gauss-Bonnet black hole is obtained as a static solution of the effective equations. The embedding in the spherical coordinates is given by
\beq\label{embedding}
H(z)=\text{sin}z, \qquad A_0=\sqrt{1-\frac{1}{L^2}}.
\eeq
In order to keep $A_0$ non-negative we demand $L\geq1$. Then the de Sitter charged GB black hole is given by a static solution
\beq\label{staticsol}
p_z(v,z)=0, \quad q(v,z)=Q, \quad m(v,z)=1-\frac{1}{L^2}+\tilde{\alpha}+Q^2.
\eeq
Here we just set the horizon radius to be unity $\sR_+=1$ for convenience, so (\ref{staticsol}) is related to (\ref{exact}) by
\beq
Q^2=\tilde{Q}^2\,r_+^{2n},\quad 1-\frac{1}{L^2}+\tilde{\alpha}+Q^2=r_+^n(1+\frac{\tilde{\alpha}}{r_+^2}),
\eeq
and  $K$ in (\ref{pq}) is related to $Q$ by
\beq
K=\frac{Q}{1-\frac{1}{L^2}+\tilde{\alpha}+Q^2}.
\eeq
Besides, as the radius $\sR$ must be positive, the physical solution requires that
\beq
Q^2 \leq 1-\frac{1}{L^2}+\tilde{\alpha},  \label{Q2}
\eeq
where the extremal case  $\sR_+=\sR_-$  corresponds to the saturated case $Q^2=1-\frac{1}{L^2}+\tilde{\alpha}$.

Now consider the perturbations  around the static solution (\ref{staticsol}) with the ansatz
\beq\label{pertur}
m(v,z)=(1-\frac{1}{L^2}+\tilde{\alpha}+Q^2)(1+\epsilon e^{-i\omega v} F_m(z)),\quad q(v,z)=Q(1+\epsilon e^{-i\omega v}F_q(z) ),
\eeq
and
\beq\label{pz}
p_z(v,z)=\epsilon e^{-i\omega v} F_z(z).
\eeq
In order to make the above expressions simpler, let us introduce
\beq\label{reducedquantity}
\overline{m}(v,z)=\frac{m(v,z)}{A_0^2},\quad \overline{q}(v,z)=\frac{q(v,z)}{A_0},
\quad \overline{Q}=\frac{Q}{A_0},\quad \overline{\alpha}=\frac{\tilde{\alpha}}{A_0^2},
\eeq
then the condition (\ref{Q2}) becomes
\beq
\overline{Q}^2\leq 1+ \overline{\alpha}, \label{Qbar}
\eeq
and eq. (\ref{pertur}) becomes
\beq\label{mandq}
\overline{m}(v,z)=(1+\overline{\alpha}+\overline{Q}^2)(1+\epsilon e^{-i\omega v} F_m(z),\quad  \overline{q}(v,z)=\overline{Q}(1+\epsilon e^{-i\omega v}F_q(z) ).
\eeq

We have two kinds of the perturbations, the charge perturbation and the gravitational
perturbation. The charge perturbation is defined by $F_m(z)\neq F_q(z)$, which describes the
fluctuation with a net charge. The gravitational perturbations which describe density fluctuation are uniquely decomposed into  scalar,  vector and tensor type. They  can be expanded in terms of the harmonic functions of different types. For the most interesting scalar-type perturbation $F_m(z)=F_q(z)$,
 which is related to the spherical harmonics $\mathbb{Y}_\ell$ on $S^{n+1}$. At a large $n$, $z=\pi/2$ we know\footnote{
This can be seen from the definition of the spherical harmonics $\Delta_{S^{n+1}}\mathbb{Y}_\ell=-\ell(\ell+n) \mathbb{Y}_\ell$ where $\Delta_{S^{n+1}}=\text{sin}^{2-n}z\frac{\partial}{\partial z}\text{sin}^{n-2}z\frac{\partial}{\partial z}+\text{sin}^{-2}z\Delta_{S^{n}}$.  At $z=\pi/2$ and in the large $n$ limit, it is easy to get that $\mathbb{Y}_\ell\sim \text{cos}^\ell z$
.} $\mathbb{Y}_\ell\sim \text{cos}^\ell z$.

To read the quasinormal modes of the perturbations, we need to impose appropriate boundary conditions on the perturbations.
The perturbations should satisfy the outgoing boundary condition at the asymptotical infinity and the ingoing boundary condition at the horizon. The latter condition has naturally been achieved as we have taken the ingoing Eddington-Finkelstein coordinates. The former condition requires that at large $\sR$
\beq \delta g_{\mu\nu}=\mc O(\sR^{-1}), \eeq
which obviously holds here.
 So plugging (\ref{pz}) and (\ref{mandq}) into the  effective equations (\ref{eff1}), (\ref{eff2}) and (\ref{eff3}), we obtain the quasinormal mode  for the charge perturbation
 \beq
 \omega_c=-i\ell,
 \eeq
so the charge perturbation is stable. This result is the same as the case without a Gauss-Bonnet term \cite{BMMT, Tanabe15}
For the scalar-type gravitational perturbation we obtain the quasinormal mode condition as
\begin{small}
\beqa\label{qnm}
&\,&\frac{2(\ell+L^2(\omega^2-\ell+i\omega\ell))(1+\overline{Q}^2+\overline{\alpha})}{L^2}+\frac{1}{L^4+2L^2(L^2-1)\overline{\alpha}}\bigg(
8i(\omega+i\ell)(-1+\ell)\overline{\alpha}(-1+\overline{Q}^2+\overline{\alpha})\nonumber\\
&\,&+2iL^4(\omega+i\ell)\big(-2-4\overline{\alpha}(1+\overline{Q}^2+\overline{\alpha})+\ell(1+\overline{\alpha}+2\overline{\alpha}^2+\overline{Q}(-1+2\overline{Q}))\big)\nonumber\\
&\,&+4L^2(-i\omega+\ell)\overline{\alpha}(-\ell+\overline{Q}^2(-4+3\ell)-4\overline{\alpha}+3\ell\overline{\alpha})\biggl)=0.
\eeqa
\end{small}
This is a quadratic equation in $\omega$, which can be solved analytically. One important feature of the above equation is that the linear term in $\omega$ is purely imaginary, so the solution is of the form
\beq
\omega_\pm =-i\omega_i \pm \omega_r
\eeq
where $\omega_i$ is a positive real number, and $\omega_r$ is the square root of something $\omega_r=\sqrt{W}$. More precisely, the solution is
\beqa\label{completeQNM}
\omega_{\pm}&=&\frac{1}{L^2(1+\overline{Q}^2+\overline{\alpha})\Big(L^2+2\overline{\alpha}(L^2-1)\Big)}\Bigg[\nonumber\\
&&-i(\ell-1)\Big(L^4+2(L^2-1)\big(1-\overline{Q}^2+L^2(1+\overline{Q}^2)\big)\overline{\alpha}+2(L^2-1)^2\overline{\alpha}^2\Big)\pm \sqrt{W(L,\overline{Q},\overline{\alpha})}\Bigg],\nonumber\\
\eeqa
where
\beqa
W(L,\overline{Q},\overline{\alpha})&=&-(\ell-1)^2\Big(L^4+2(L^2-1)\big(1-\overline{Q}^2+L^2(1+\overline{Q}^2)\big)\overline{\alpha}+2(L^2-1)^2\overline{\alpha}^2\Big)^2\nonumber\\
&&+L^2\ell(1+\overline{Q}^2+\overline{\alpha})\Big(L^2+2(L^2-1)\overline{\alpha}\Big)\Bigg(2\Big(3+\overline{Q}^2(2\ell-1))+2\ell(\overline{\alpha}-1)-\overline{\alpha}\Big)\overline{\alpha}
\nonumber\\
&&-L^2\Big(1+\overline{Q}^2(1-4\overline{\alpha}+6\ell\overline{\alpha})+\overline{\alpha}(5-4\overline{\alpha}+\ell(6\overline{\alpha}-2)\overline{\alpha}\Big)\nonumber\\
&&+L^4(\ell-1)\Big(1+\overline{\alpha}+2\overline{\alpha}^2+\overline{Q}^2(2\overline{\alpha}-1)\Big)
\Bigg).
\eeqa
 In the limit $\overline{\alpha}\to0$ the solutions reproduce the ones for the de Sitter Reissner-Nordstrom black hole in \cite{Tanabe15}.  When $W<0$, the  frequency of the quasinormal mode becomes purely imaginary, and especially $\omega_+=i(\sqrt{-W}-\omega_i)$ may have a positive imaginary part, suggesting the black hole is unstable. The necessary condition that $\omega_+$ has a positive imaginary part is
\beqa\label{completeNC}
&&2\Big(3+\overline{Q}^2(2\ell-1))+2\ell(\overline{\alpha}-1)-\overline{\alpha}\Big)\overline{\alpha}+L^4(\ell-1)\Big(1+\overline{\alpha}+2\overline{\alpha}^2+\overline{Q}^2(2\overline{\alpha}-1)\Big)
\nonumber\\&&-L^2\Big(1+\overline{Q}^2(1-4\overline{\alpha}+6\ell\overline{\alpha})+\overline{\alpha}(5-4\overline{\alpha}+\ell(6\overline{\alpha}-2)\overline{\alpha}\Big)<0.
\eeqa

\paragraph{Asymptotically flat Gauss-Bonnet black hole}

Firstly, let us consider the asymptotically flat case which corresponds to $L\to \infty $. For the neutral case $\overline{Q}=0$, we easily obtain the frequencies  of the scalar-type  quasinormal mode as
\beq
\omega_{\pm}=\frac{-i(\ell-1)(1+2\overline{\alpha}+2\overline{\alpha}^2)\pm\sqrt{(\ell-1)\Big[\Big(1+2\overline{\alpha}(1+\overline{\alpha})\Big)^2
-\ell \overline{\alpha}^2\Big]}}{(1+\overline{\alpha})(1+2\overline{\alpha})}.
\eeq
In the small and large $\overline{\alpha}$ limits, we get
\beq
\omega_{\pm}=\pm\sqrt{\ell-1}+i(\ell-1),
\eeq
which agree exactly with the ones found in \cite{CFLY}, suggesting the black hole is stable. For a generic Gauss-Bonnet parameter, it  has not been discussed in \cite{CFLY}, as the analysis based on the master equations becomes rather difficult. Here it is fairly easy to read the quasinormal modes. In this case the necessary condition (\ref{completeNC}) becomes
 \beq\label{AFunchargedGB}
(\ell-1)(1+\overline{\alpha}+2\overline{\alpha}^2)<0,
 \eeq
which can never be satisfied. It turns out that there is no instability for the neutral Gauss-Bonnet black hole at large $D$ for any value of the Gauss-Bonnet coefficient.
 The result is consistent with the one found in \cite{Konoplya} that when the spacetime dimension is  large enough, the Gauss-Bonnet black hole is always dynamically stable.

When the charge is taken into account,  from (\ref{completeQNM}) we find that
\begin{small}
\beq\label{chargedGB}
\omega_{\pm}=\frac{-i(\ell-1)\Big(1+2(1+\overline{Q}^2)\overline{\alpha}+2\overline{\alpha}^2\Big)\pm\sqrt{(\ell-1)\Big[\Big(1+2(1+\overline{Q}^2)\overline{\alpha}
+2\overline{\alpha}^2\Big)^2-\ell(\overline{Q}^2+\overline{\alpha})^2\Big]}}{(1+2\overline{\alpha})(1+\overline{\alpha}+\overline{Q}^2)}.
\eeq
\end{small}
The necessary condition (\ref{completeNC}) now becomes
 \beq
(\ell-1)\Big(1+\overline{\alpha}+2\overline{\alpha}^2+\overline{Q}^2(2\overline{\alpha}-1)\Big)<0,
 \eeq
since the charge is restricted to $\overline{Q}^2\leq1+\overline{\alpha}$, the above condition can never be satisfied. Hence the imaginary parts of the frequencies are  always negative.
This suggests that the asymptotically flat  charged Gauss-Bonnet black hole at large $D$  is always stable under the scalar-type gravitational perturbation, similar to the case of Reissner-Nordstrom black hole \cite{BMMT}, although until now there is no numerical study to verify this. In addition, it would be interesting to note that at extremity, that is  $\overline{Q}^2=1+\overline{\alpha}$, from (\ref{chargedGB}) we find that $\omega_+$ becomes
\beq
\omega_{+}=\frac{-i(\ell-1)\Big(1+2\overline{\alpha}\Big)+\sqrt{(\ell-1)\Big[\Big(1+2\overline{\alpha}
\Big)^2-\ell\Big]}}{2(1+\overline{\alpha})}.
\eeq
For the extremal Reissner-Nordstrom black hole $\overline{\alpha}=0$ such that $\omega_+=0$. In contrast, for the extreaml Gauss-Bonnet black hole $\omega_+\neq0$.
\begin{figure}[t]
 \begin{center}
  \includegraphics[width=65mm,angle=0]{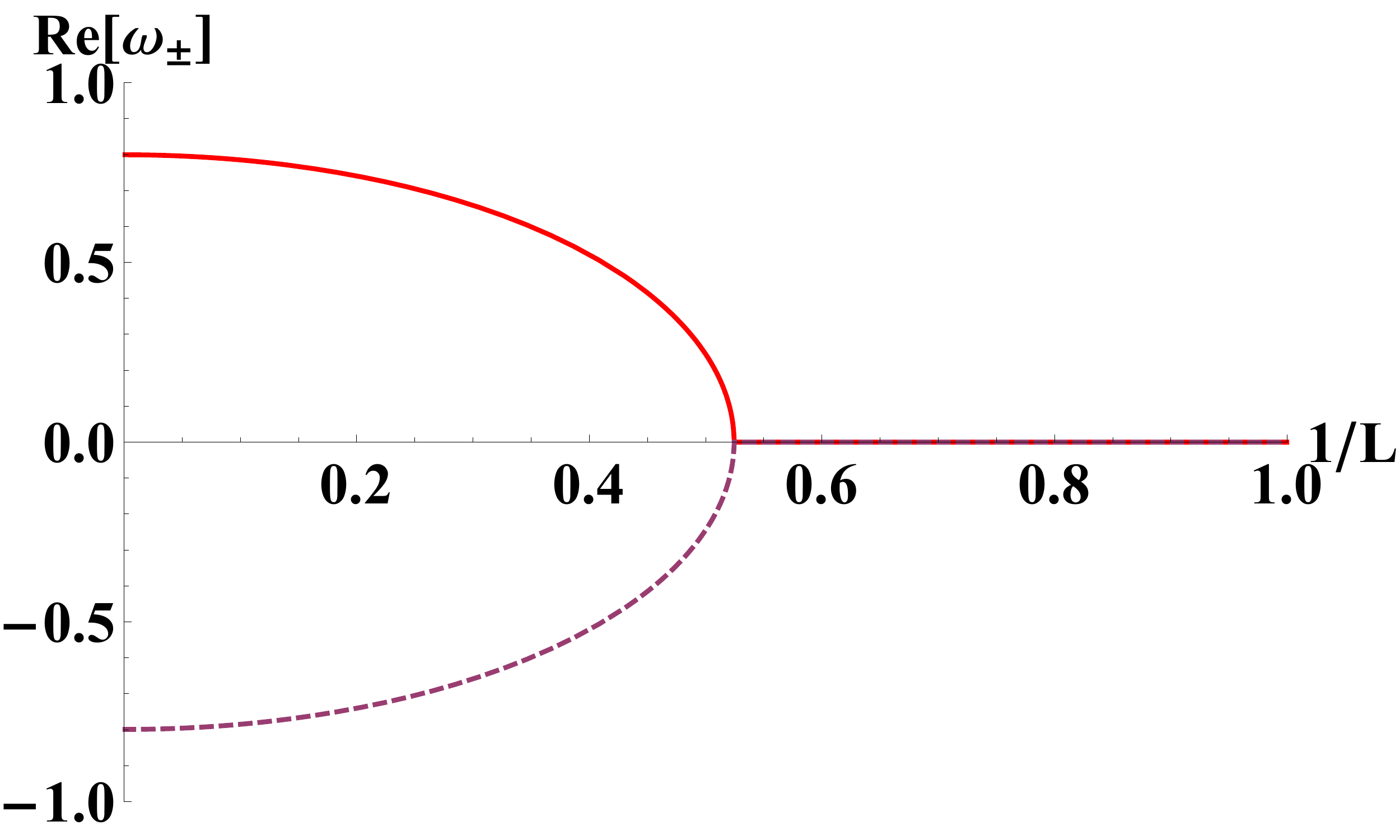}
  \hspace{5mm}
  \includegraphics[width=65mm,angle=0]{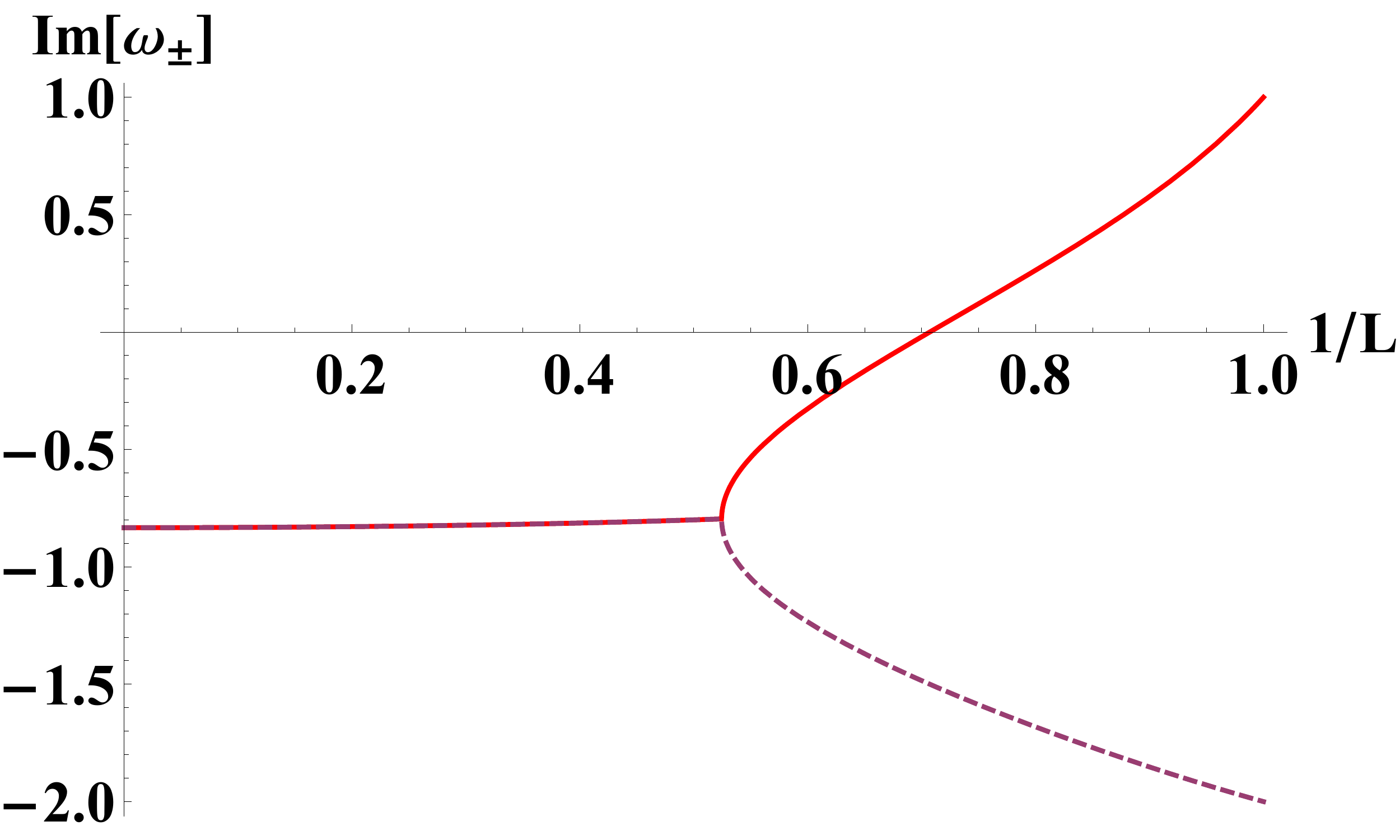}
 \end{center}
 \vspace{-5mm}
 \caption{The quasinormal mode frequencies $\omega_{+}$ (solid line) and $\omega_{-}$ (dashed line) of the gravitational perturbation of de Sitter Gauss-Bonnet black hole with $\ell=2$ and $\overline{\alpha}=1$. The real part and the imaginary part are depicted in the left and right panel respectively.
 When $1/L>1/\sqrt{2}\approx0.71$, the perturbation becomes unstable.}
 \label{fig1}
\end{figure}

\paragraph{de Sitter Gauss-Bonnet black hole}

When there is a positive cosmological constant, the stability of the Gauss-Bonnet black hole is different.   In \cite{Cuyubamba} the authors found that the Gauss-Bonnet black hole in de Sitter spacetime becomes unstable
for $D\geq5$ at sufficiently large values of the cosmological constant $\Lambda$, by using numerical analysis. This new kind of instability is called ``the $\Lambda$-instability", because it does not take place for asymptotically flat spacetime. Here we can  show  this point explicitly in the large $D$ limit.

Consider first the neutral Gauss-Bonnet black hole without charge $\overline{Q}=0$, from (\ref{completeQNM}) we obtain
\beqa\label{neutralGBQNM}
\omega_{\pm}=\frac{1}{L^2(1+\overline{\alpha})(L^2+(-1+L^2)\overline{\alpha})}\Bigg[&-&i(\ell-1)\Big(L^4+2(-1+L^4)\overline{\alpha}+2(-1+L^2)^2\overline{\alpha}^2 \Big)\nonumber\\
&\pm &\sqrt{W(L,\overline{\alpha})}\Bigg],
\eeqa
where
\beqa\label{WLalpha}
W(L,\overline{\alpha})
&=&-(\ell-1)^2\Big(L^4+2(-1+L^4)\overline{\alpha}+2(-1+L^2)^2\overline{\alpha}^2\Big)^2\nonumber\\
&\,&+L^2\ell(1+\overline{\alpha})\Big(L^2+2(-1+L^2)\overline{\alpha}\Big)
\Big(2(3+2\ell(-1+\overline{\alpha})-\overline{\alpha})\overline{\alpha}\nonumber\\
&\,&+L^4(-1+\ell)(1+\overline{\alpha}+2\overline{\alpha}^2)+L^2(-1+\overline{\alpha}(-5+\ell(2-6\overline{\alpha})+4\overline{\alpha})) \Big) .\nonumber
\eeqa
From this formula it is not hard to determine when the quasinormal mode frequency has a positive imaginary part so that the black hole becomes unstable. For example, when $L$ is sufficiently small, the term $W(L,\overline{\alpha})$ becomes negative, so the frequency $\omega_+$ may have a positive imaginary part. Actually, the necessary condition (\ref{completeNC}) that the frequency $\omega_+$  has a positive imaginary part is
\beq\label{dSGBNC}
2(3+2\ell(-1+\overline{\alpha})-\overline{\alpha})\overline{\alpha}+L^4(-1+\ell)(1+\overline{\alpha}+2\overline{\alpha}^2)+L^2(-1+\overline{\alpha}(-5+\ell(2-6\overline{\alpha})+4\overline{\alpha}))<0.
\eeq

 \begin{figure}[t]
 \begin{center}
  \includegraphics[width=55mm,angle=0]{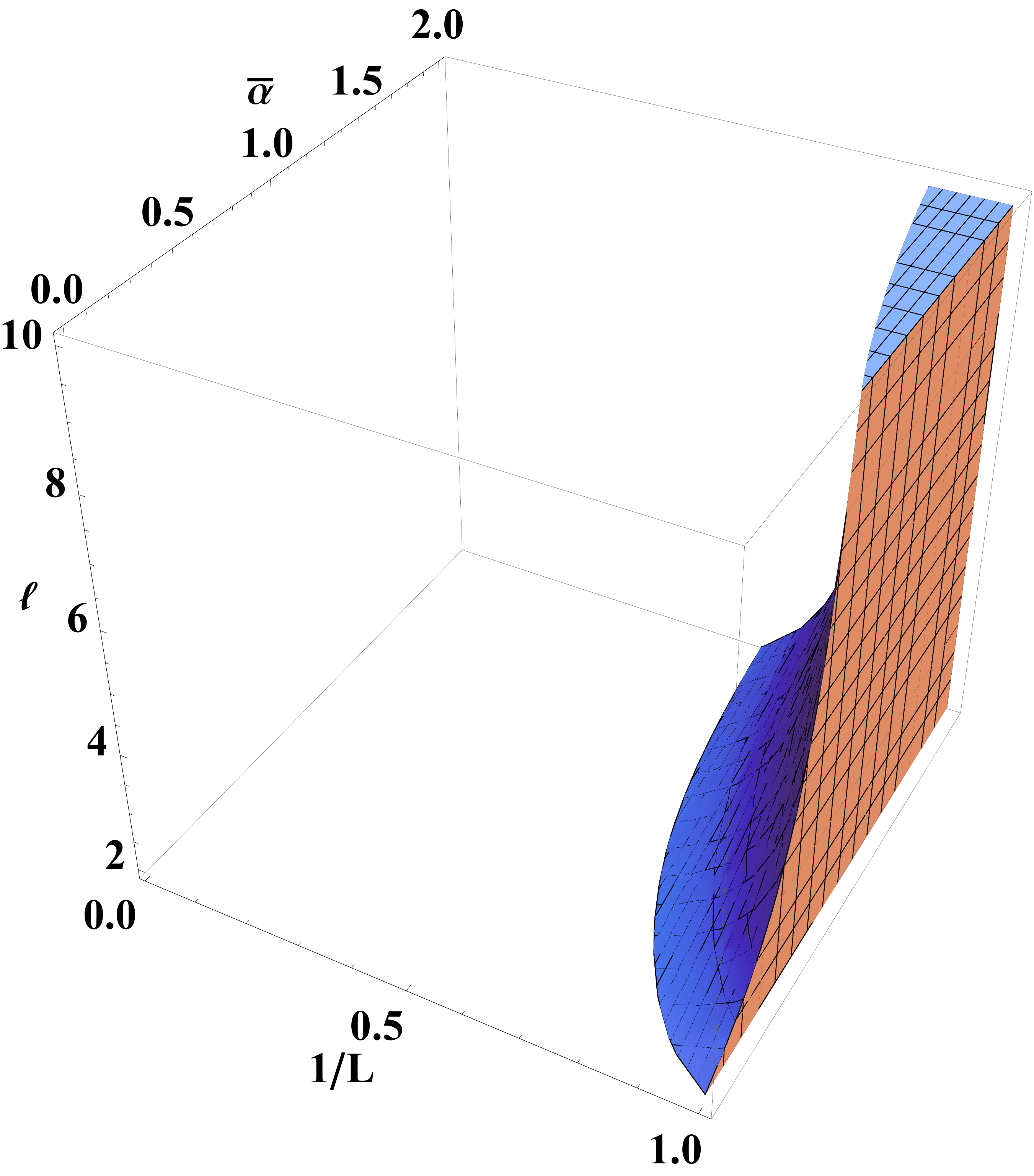}
    \hspace{25mm}
  \includegraphics[width=55mm,angle=0]{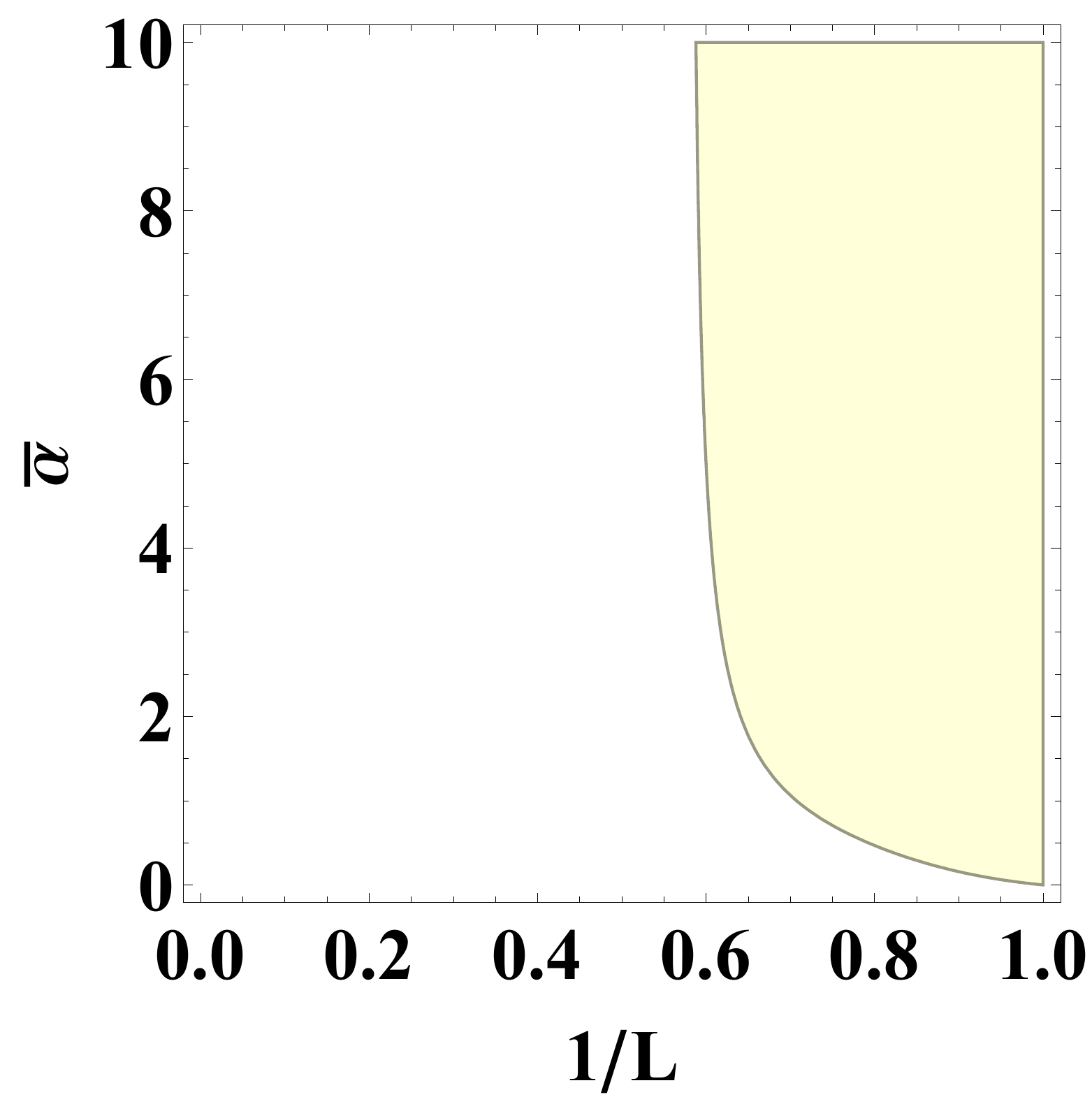}
 \end{center}
 \vspace{-5mm}
 \caption{The unstable regions of the de Sitter Gauss-Bonnet black hole. The left panel  shows the unstable region in $(L, \overline{\alpha}, \ell)$ space. Note that $\ell=2, 3,...$, here for simplicity we take $\ell$ to be continuous in this plot and in Fig. \ref{fig3}.  The right panel shows  the unstable region with  $\ell=2$  in light yellow.}
 \label{fig2}
\end{figure}

From Fig. \ref{fig1} we see that the quasinormal mode frequencies $\omega_{\pm}$  of de Sitter Gauss-Bonnet black hole become purely imaginary when $1/L>0.52$, and  $\mbox{Im}[\omega_{+}]>0$ when $1/L>1/\sqrt{2}$ suggesting the black hole could be unstable under a perturbation. In Fig. \ref{fig2} we show the unstable region of the de Sitter Gauss-Bonnet black hole.   It is easy to see that for a given $\overline{\alpha}$, when $L$ is smaller than a critical value $L_\ell$ the de Sitter Gauss-Bonnet black hole can always be unstable. This is consistent with the conclusion in  \cite{Cuyubamba} that when the cosmological constant is large enough, the black hole is unstable. From (\ref{neutralGBQNM}), in the limit $\overline{\alpha}\to \infty$, it is easy to obtain
\beq\label{criticalL}
L_\ell=\sqrt{\frac{2\ell-1}{\ell-1}}.
\eeq

\begin{figure}[t]
 \begin{center}
\includegraphics[width=55mm,angle=0]{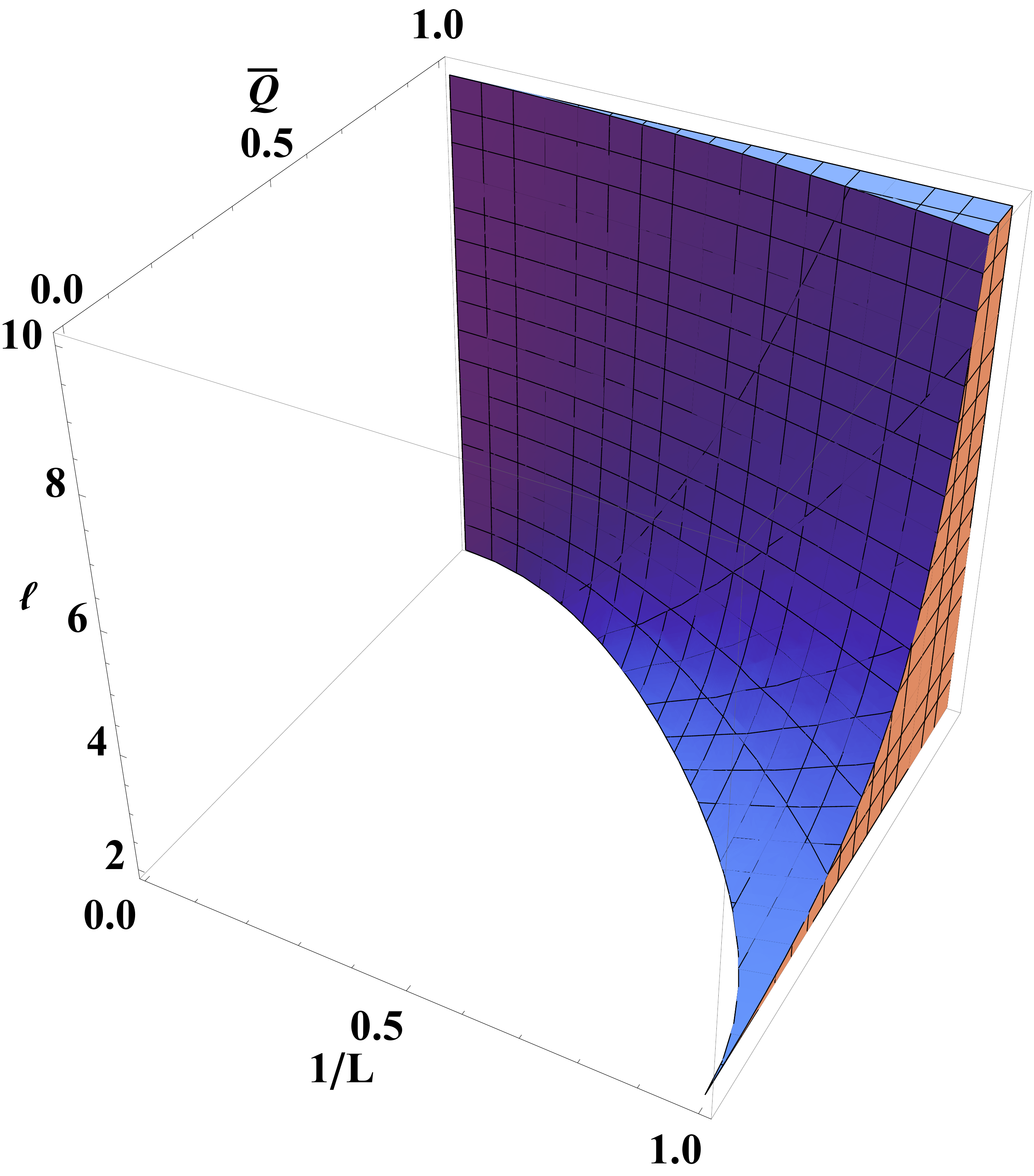}
  \hspace{25mm}
 \includegraphics[width=55mm,angle=0]{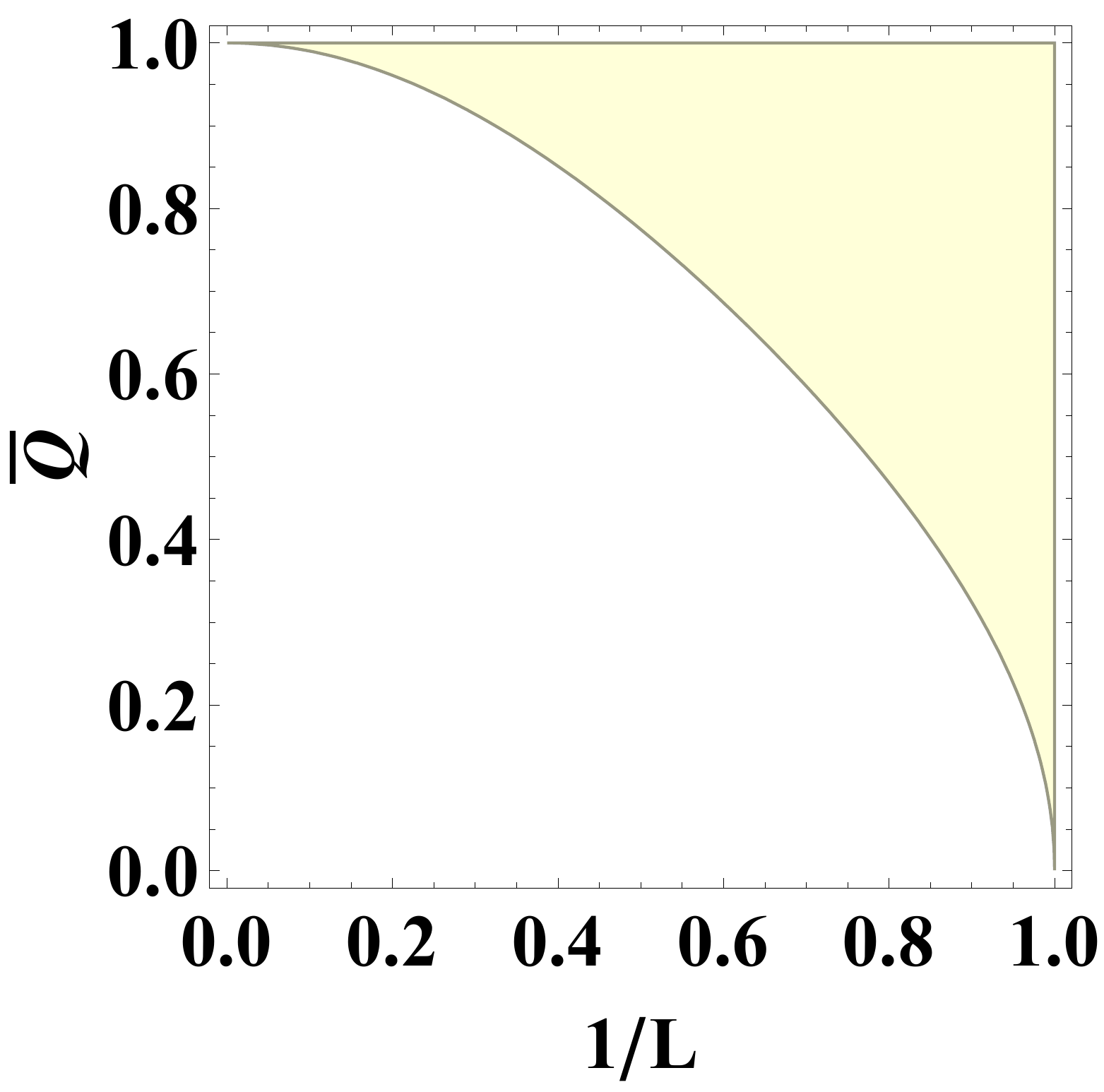}
 \end{center}
 \vspace{-5mm}
 \caption{The unstable regions of  the de Sitter Reissner-Nordstrom black hole.  The left panel  shows the unstable region in $(L, \overline{Q}, \ell)$ space.  The right panel shows  the unstable region with  $\ell=2$  in light yellow.}
 \label{fig3}
\end{figure}

It would be interesting to compare the ``$\Lambda$-instability" in the de Sitter Gauss-Bonnet black hole with the one in the de Sitter Reissner-Nordstrom black hole.  In the latter case, for a fixed charge $\overline{Q}$, the instability appears when the cosmological constant obeys the following relation
\beq
\frac{1}{L^2}>\frac{1}{L_c^2}=\frac{1-\overline{Q}^2}{1+\overline{Q}^2}(\ell-1).
\eeq
Obviously the presence of the charge can lower the critical value of the cosmological constant. On the other hand, for a fixed cosmological constant, when the charge is larger than the critical value \cite{Tanabe15}
\beq\label{criticalcharge}
\overline{Q}_\ell=\sqrt{\frac{(\ell-1)-1/L^2}{(\ell-1)+1/L^2}},
\eeq
the de Sitter Reissner-Nordstrom black hole becomes unstable.  In Fig. \ref{fig3} we show the unstable region in the de Sitter Reissner-Nordstrom black hole. From Fig. \ref{fig2} and Fig. \ref{fig3} we can see that in both cases larger $\ell$'s reduce the size of the unstable region, so $\ell=2$ corresponds to the most unstable mode.

Note that from the left panel of Fig. \ref{fig2}, we can see that the unstable region of the de Sitter Gauss-Bonnet black hole seems to extend to $L=1$ and $\overline{\alpha}=0$.
However, $L=1$ leads to $A_0=0$, the form (\ref{reducedquantity}) is not correct any more. Instead we should use the original quantities (\ref{pertur}) with $Q=0$ and $\tilde{\alpha}=0$. This obviously means that our solution breaks down in this limit since $m(v, z)=0$. This  occurs also for the de Sitter Reissner-Nordstrom black hole \cite{Tanabe15} as observed in the left panel of Fig. \ref{fig3}. For the de Sitter Reissner-Nordstrom black hole $L=1$ is  the Nariai limit but here the Nariai limit happens at $L=1/\sqrt{1+\tilde{\alpha}}$ which is beyond the range we can touch as we require $L\ge1$.

\begin{figure}[t]
 \begin{center}
  \includegraphics[width=75mm,angle=0]{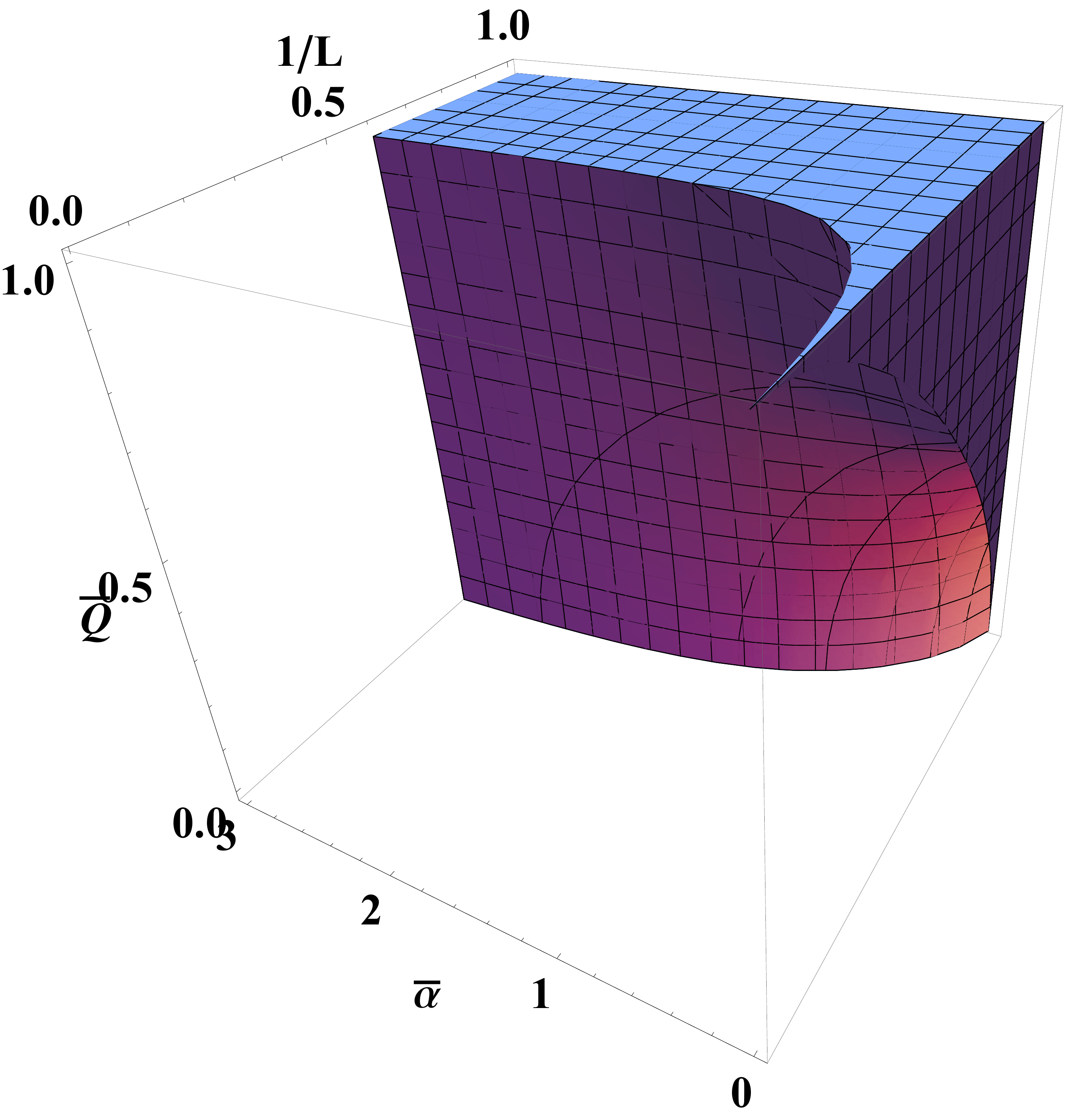}
  \end{center}
 \vspace{-5mm}
 \caption{The unstable region of de Sitter charged Gauss-Bonnet black hole with $\ell=2$ is shown in $(\overline{\alpha}, \overline{Q}, L)$ space.}
 \label{fig4}
\end{figure}

\begin{figure}[t]
 \begin{center}
  \includegraphics[width=55mm,angle=0]{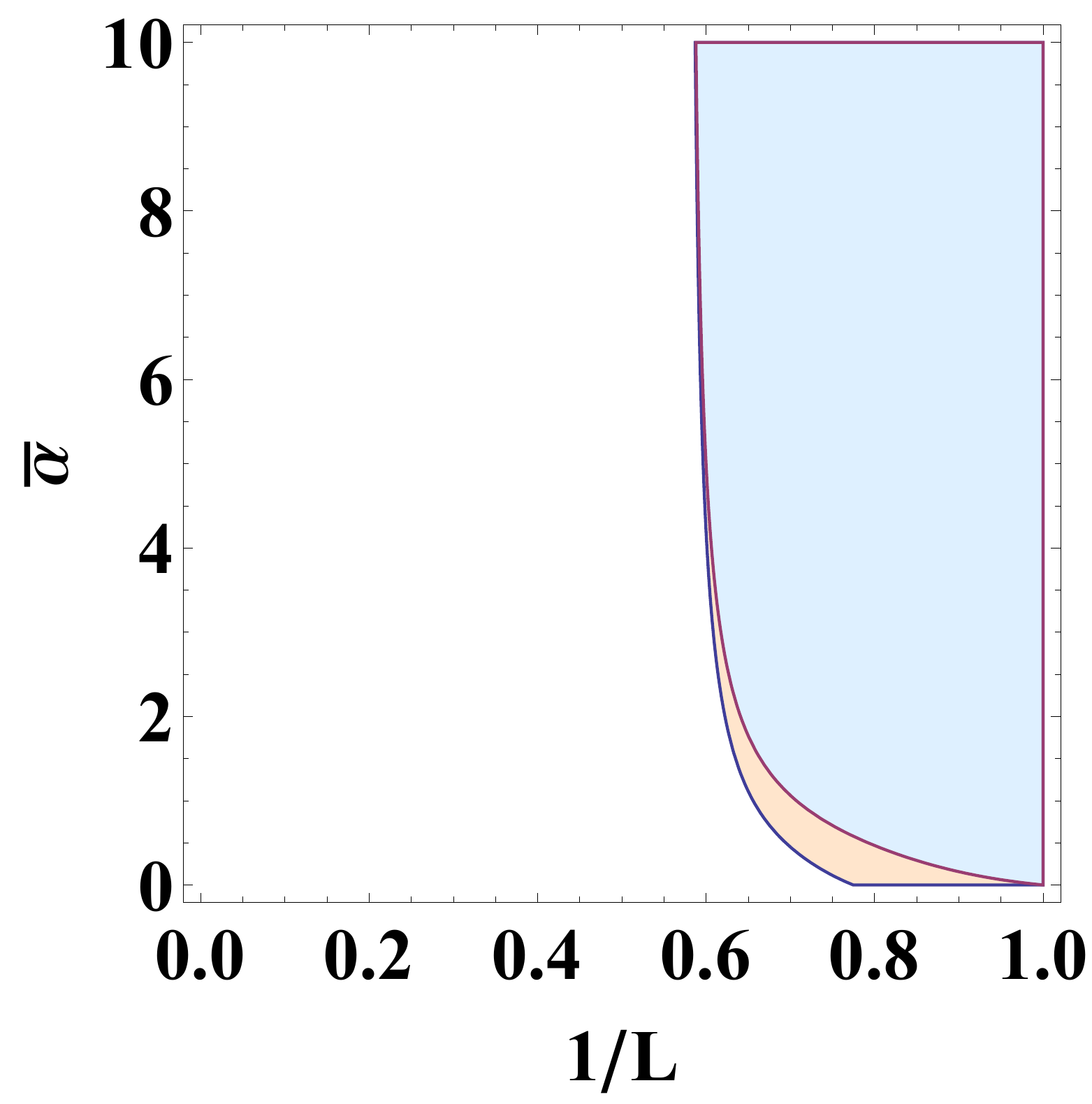}
 \hspace{25mm}
  \includegraphics[width=55mm,angle=0]{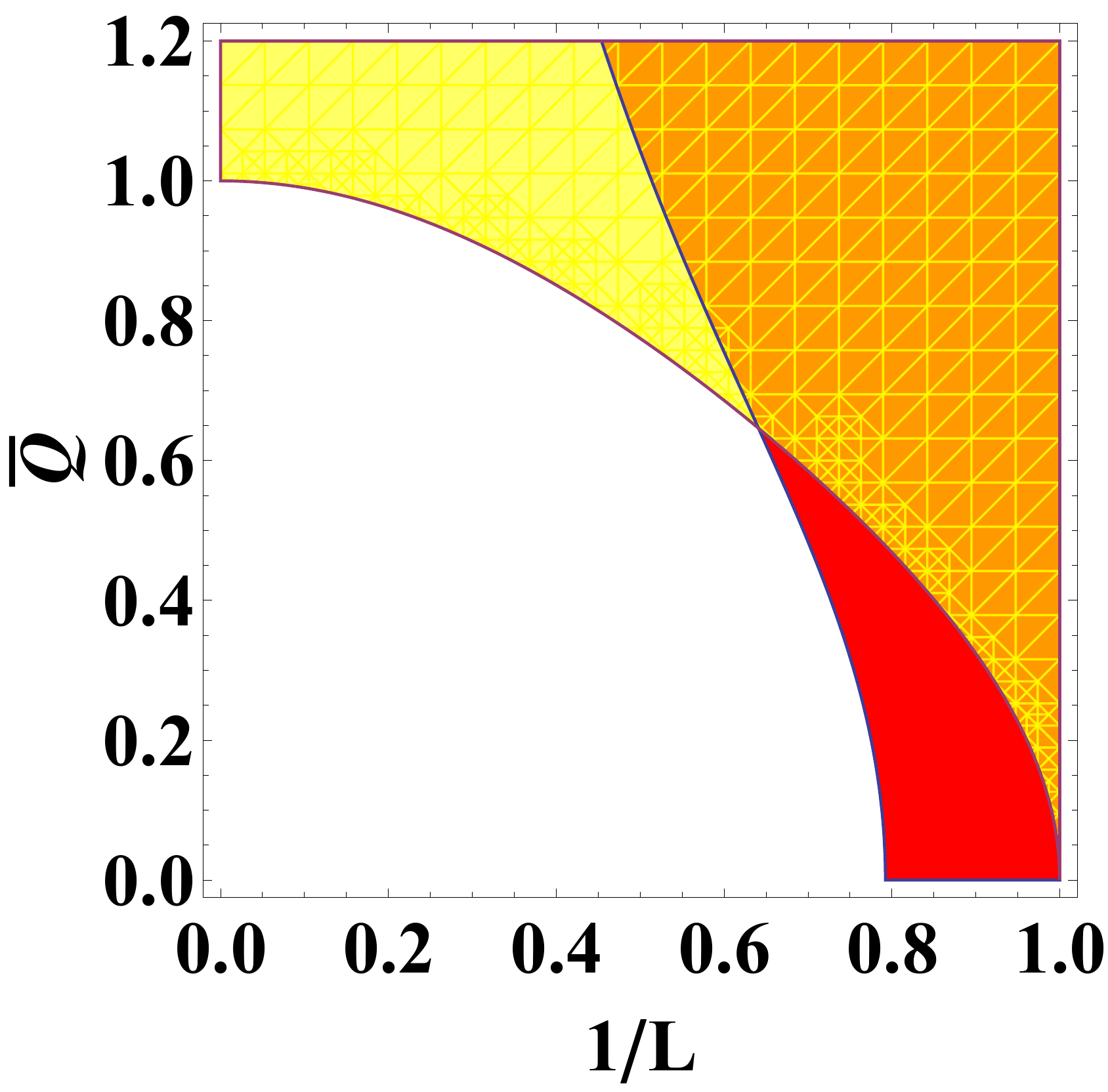}
 \end{center}
 \vspace{-5mm}
 \caption{The unstable region of the de Sitter charged Gauss-Bonnet black hole with $\ell=2$. The left panel  shows the unstable region in $(L, \overline{\alpha})$ plane, where the light blue region corresponds to $\overline{Q}=0$ and the light blue region plus the light orange region correspond
 to $\overline{Q}=0.5$. The right panel  shows the unstable region in $(L, \overline{Q})$ plane,  where the yellow region plus the orange region correspond to $\overline{\alpha}=0$ and  the red region plus the orange region correspond
 to $\overline{\alpha}=0.5$. }
 \label{fig5}
\end{figure}

\begin{figure}[t]
 \begin{center}
  \includegraphics[width=35mm,angle=0]{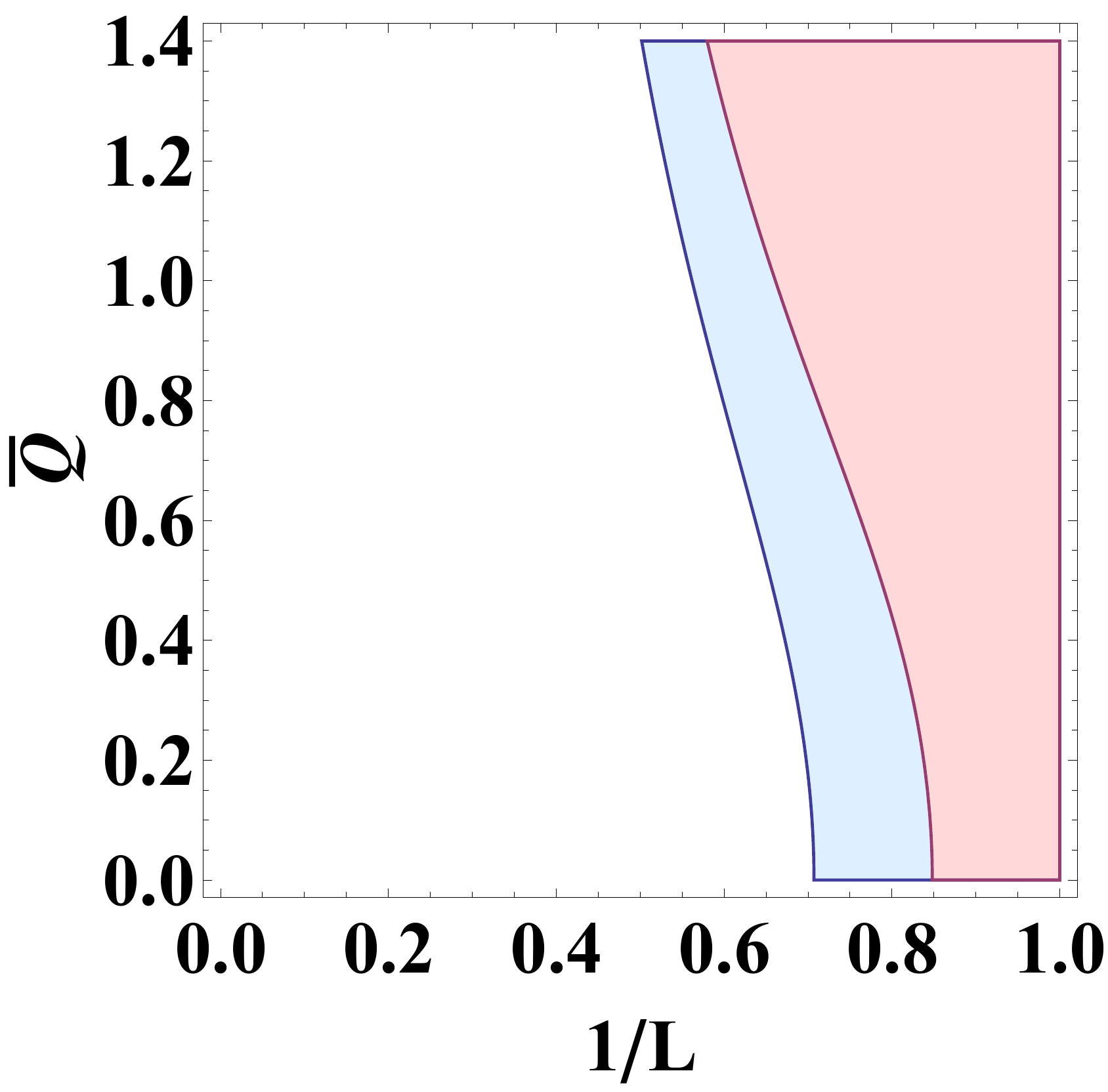}
 \hspace{1mm}
  \includegraphics[width=35mm,angle=0]{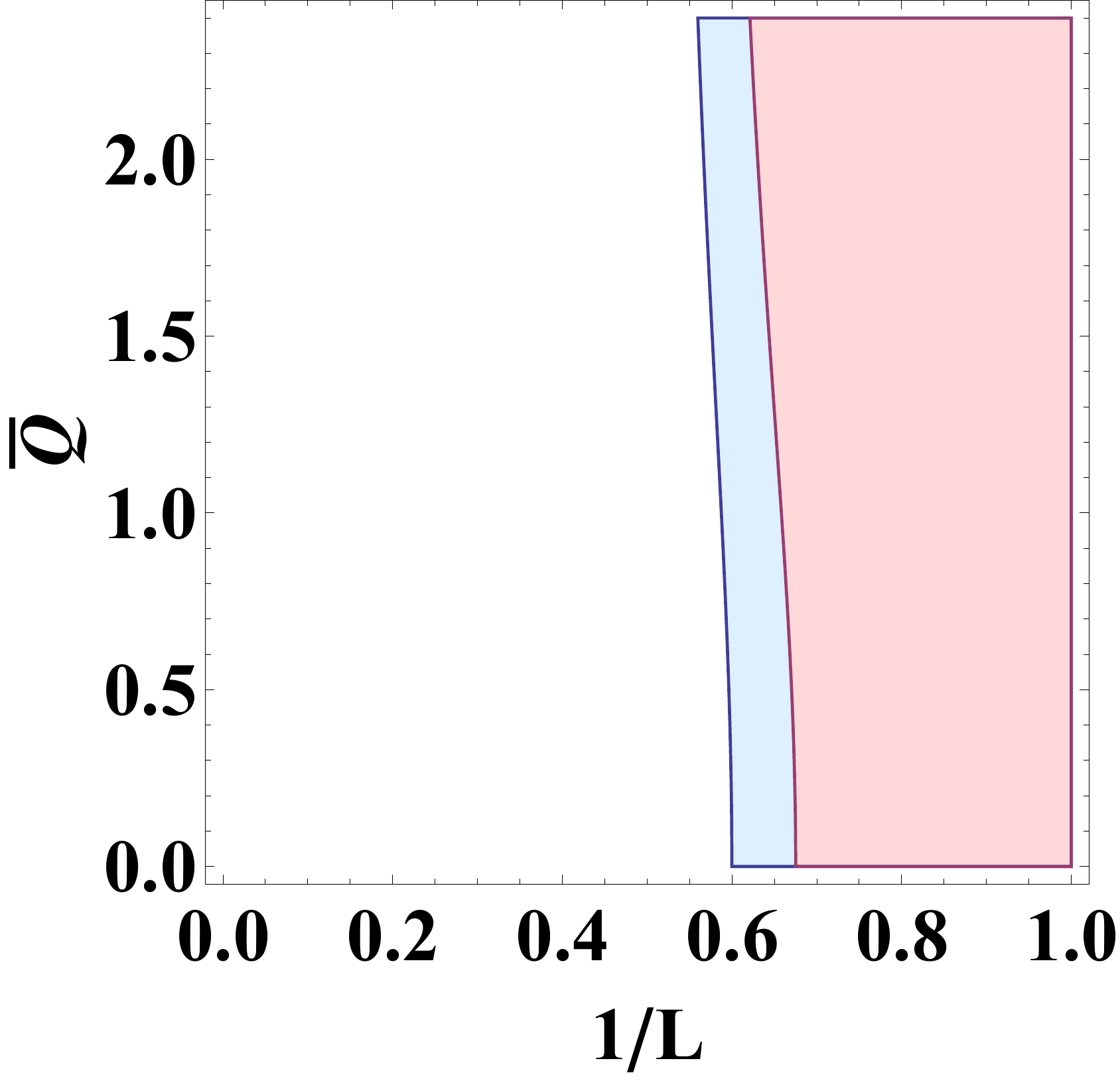}
   \hspace{1mm}
  \includegraphics[width=35mm,angle=0]{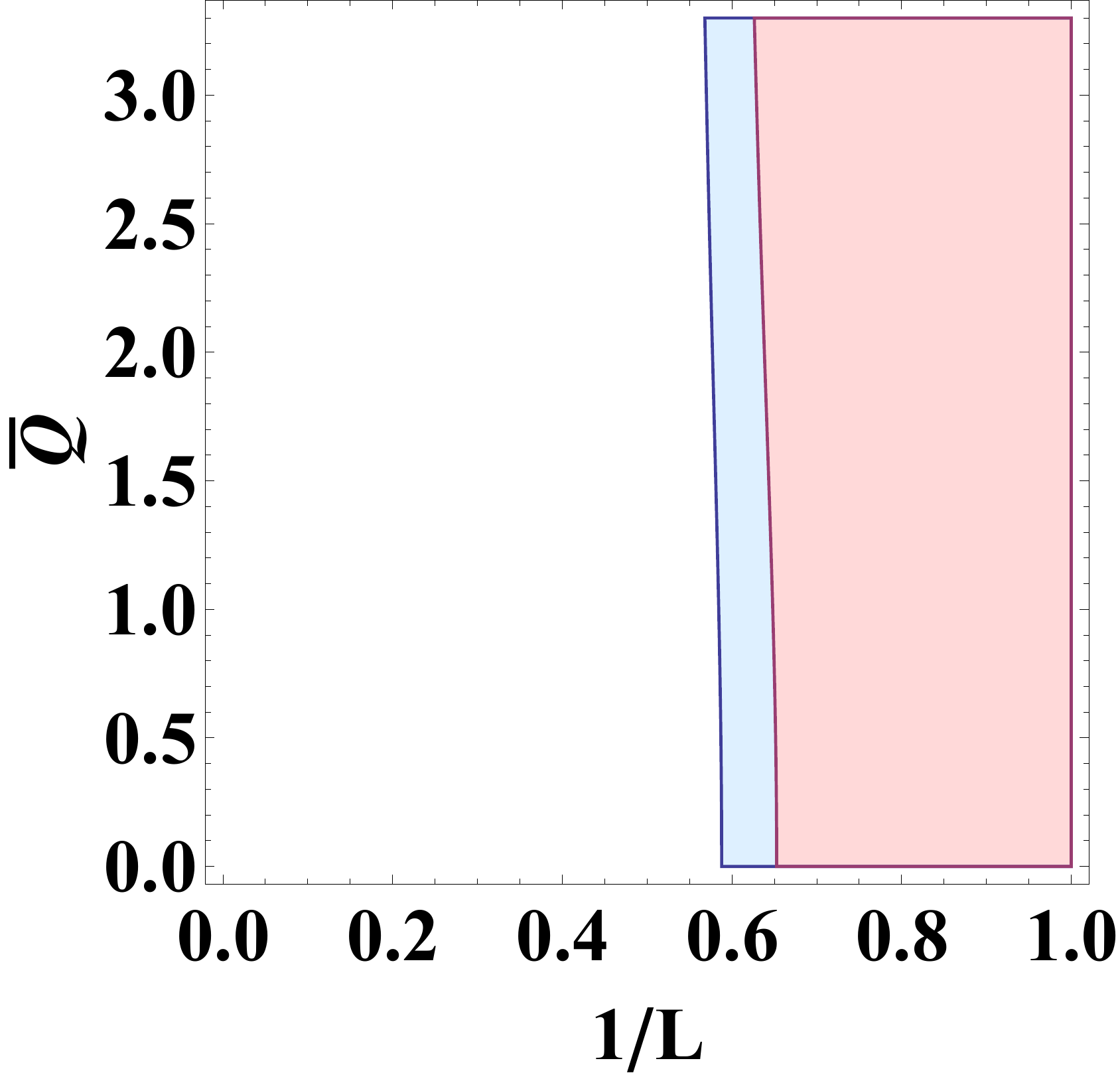}
  \hspace{1mm}
  \includegraphics[width=35mm,angle=0]{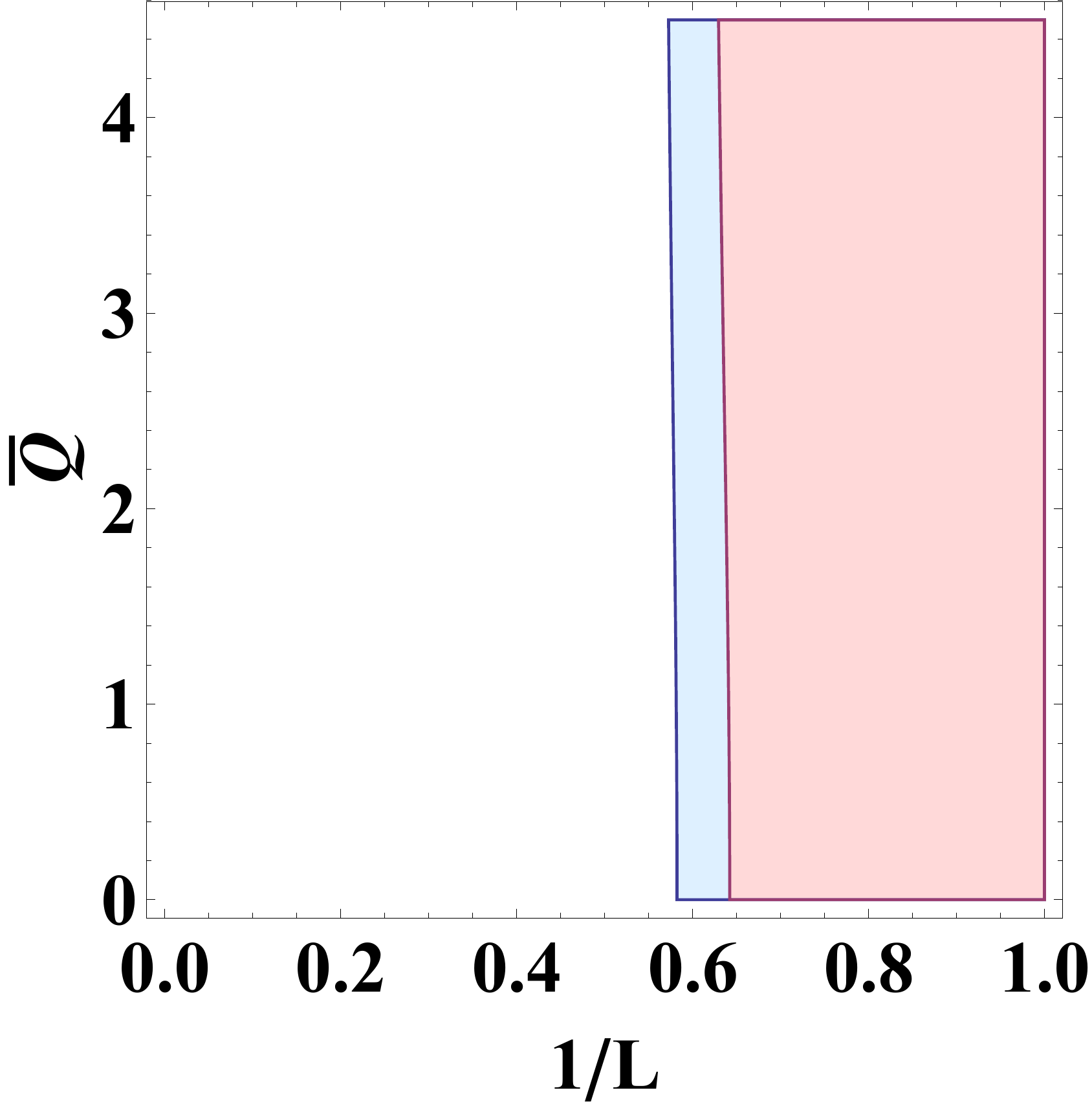}
 \end{center}
 \vspace{-5mm}
 \caption{The unstable region of the de Sitter charged Gauss-Bonnet black hole. From the left to the right $\overline{\alpha}=$ 1, 5, 10 and 20, respectively, where the light red region corresponds to $\ell=3$ and the light red region plus the light blue region correspond to $\ell=2$.}
 \label{fig6}
\end{figure}
When both the charge and the  Gauss-Bonnet term are taken into account, the instability of the black hole becomes complex. For simplicity, we take $\ell=2$ to illustrate their effects on the instability. In Fig. \ref{fig4} we depict the unstable region of  the de Sitter charged  Gauss-Bonnet black hole in $(\overline{\alpha}, \overline{Q}, L)$ space. In order to show the effect of the charge, we compare the unstable regions with and without the charge in the left panel of Fig. \ref{fig5}.  We can see that the presence of  the charge enlarges the unstable region and lowers the critical value $L_{\ell}$. To show the effect of the Gauss-Bonnet term, we compare the unstable regions with and without the term in  the right panel  of Fig. \ref{fig5}.  The presence of the  Gauss-Bonnet coefficient $\overline{\alpha}$ enlarges the range of $\overline{Q}$ due to the relation $\overline{Q}^2\leq1+\overline{\alpha}$, but the effect of $\overline{\alpha}$ on the instability of the black hole depends on the value of the cosmological constant. When $L$ is large, $\overline{\alpha}$ helps to stabilize the black hole, such that in some range of $L$ the black hole is always stable no matter what value the charge is  taken.
However, when $L$ is small, $\overline{\alpha}$ helps  to destabilize the black hole. Moreover, if $\overline{\alpha}$ is very large, the stability of the black hole is  totally determined by  the cosmological constant, $L>L_\ell$ or $L<L_\ell$,  but being  independent of the charge, which can be seen in Fig. \ref{fig6}. In addition, the angular quantum number $\ell$ plays the same role as in the case without the Gauss-Bonnet term or the charge, so $\ell=2$ corresponds to the most unstable mode of the de Sitter charged  Gauss-Bonnet black hole.

\paragraph{AdS charged Gauss-Bonnet black hole} The above discussion can be easily applied to the AdS case by replacing $L\to iL$, except that in this case we require $4\tilde{\alpha}/L^2\leq1$, or in terms of $\overline{\alpha}$, which is expressed as
\beq
\overline{\alpha}\leq\frac{L^4}{4(1+L^2)},
\eeq
in order to have a well-defined vacuum.
The analogous effective equations and quasinormal modes can be  obtained after such a replacement. From (\ref{completeNC}), after the replacement $L\to iL$ we obtain the necessary condition for the existence of
unstable mode
\beqa
&&L^4(\ell-1)\Big(1+\overline{\alpha}+2\overline{\alpha}^2+\overline{Q}^2(2\overline{\alpha}-1)\Big)+L^2\Big(1+\overline{Q}^2(1-4\overline{\alpha}+6\ell\overline{\alpha})+\overline{\alpha}(5-4\overline{\alpha}+\ell(6\overline{\alpha}-2)\overline{\alpha}\Big)
\nonumber\\&&+2\Big(3+\overline{Q}^2(2\ell-1))+2\ell(\overline{\alpha}-1)-\overline{\alpha}\Big)\overline{\alpha}<0.
\eeqa
Then it is evident that the AdS charged Gauss-Bonnet black hole is always stable at large $D$.

\paragraph{Deformed static solution}

It was found in \cite{Konoplya:2008au} that for  the de Sitter charged black hole in higher dimensions there exists deformed black hole solution at the edge of the instability. This kind of deformed solution is not spherically symmetric any more. In \cite{Tanabe15}, the deformed solution of the de Sitter charged black hole has been constructed from the large $D$ effective equations. It is interesting to see if there is similar phenomenon  for  the Gauss-Bonnet gravity.

From the perturbation analysis we notice that the deformed static solution does exist in the Gauss-Bonnet gravity.  Since the unstable mode is always of a purely imaginary frequency, at the edge of  the instability, there exists a non-trivial zero-mode ($\omega=0$) static perturbation, suggesting the existence of a non-spherical symmetric static  solution branch. By solving the large $D$ effective equations (\ref{eff1}), (\ref{eff2}) and (\ref{eff3}) in the static case, we are able to find such a static non-spherical symmetric de Sitter charged Gauss-Bonnet black hole solution.
In the background of  de Sitter spacetime (\ref{embedding}), from (\ref{pq}) and (\ref{Pz}) in terms of $\overline{m}(z)$, $\overline{q}(z)$, $p_z(z)$ and $\overline{Q}$, the static solution is given by
\beq\label{deformedsolutionpzqm}
p_z(z)=\overline{m}'(z),\quad \overline{q}(z)=\frac{\overline{Q}}{1+\overline{\alpha}+\overline{Q}^2} \overline{m}(z),\quad \overline{m}(z)=e^{P(z)},
\eeq
where
\beq\label{deformedsolutionPz}
P(z)=P_0+P_1(\text{cos}~z)^{P_2},
\eeq
\beq\label{deformedsolutionP2}
P_2=\frac{2\overline{\alpha}(-3+\overline{\alpha}+\overline{Q}^2)+L^2(1+\overline{Q}^2(1-4\overline{\alpha})+(5-4\overline{\alpha})\overline{\alpha})+L^4(1+\overline{\alpha}+2\overline{\alpha}^2+\overline{Q}^2(-1+2\overline{\alpha}))}
{4\overline{\alpha}(-1+\overline{Q}^2+\overline{\alpha})-2L^2\overline{\alpha}(-1+3\overline{Q}^2+3\overline{\alpha})+L^4(1+\overline{\alpha}+2\overline{\alpha}^2+\overline{Q}^2(-1+2\overline{\alpha}))}.
\eeq
$P_0$ and $P_1$ are the integration constants, whose physical meanings become clear after comparing with (\ref{mandq}).  If we set the horizon radius $\sR_+=1$ then $e^{P_0}=1+\overline{\alpha}+\overline{Q}^2$, so $P_0$ is an $\mc O(1/n)$ redefinition of the horizon radius. $P_1$  describes an $\mc O(1/n)$  amplitude deformed from spherical symmetry. The solutions is not analytic at $z=\pi/2$ for general $\overline{Q},\overline{\alpha}$ and $1/L$. However, at the edge of the instability, the right hand side of (\ref{completeNC}) must be zero such that $P_2=\ell$. In this case, the solution becomes regular. For $\ell \geq 2$, the solution is a static solution without spherical symmetry.  

\section{The case $\tilde{\alpha}<0$}\label{appendixA}

Now we discussion the case that the Gauss-Bonnet term has a negative coefficient, i.e.
$\tilde{\alpha}<0$. From (\ref{R+}) we find a restriction on the value of $\overline{\alpha}$, that is
\beq\label{restriction}
\overline{\alpha}>-1,
\eeq
otherwise the solution is not well-defined.  With this condition it is easy to extend the previous discussion to the case $\overline{\alpha}<0$.  Here we focus on the stabilities of the solutions. The necessary condition for the scalar gravitational perturbation  to have an unstable
mode is given by
\beqa\label{minusalphaNC}
&&\Big(L^4+2L^2(L^2-1)\overline{\alpha}\Big)\Bigg(2\Big(3+\overline{Q}^2(2\ell-1))+2\ell(\overline{\alpha}-1)-\overline{\alpha}\Big)\overline{\alpha}
\nonumber\\
&&-L^2\Big(1+\overline{Q}^2(1-4\overline{\alpha}+6\ell\overline{\alpha})+\overline{\alpha}(5-4\overline{\alpha}+\ell(6\overline{\alpha}-2)\overline{\alpha}\Big)\nonumber\\
&&+L^4(\ell-1)\Big(1+\overline{\alpha}+2\overline{\alpha}^2+\overline{Q}^2(2\overline{\alpha}-1)\Big)
\Bigg)<0.
\eeqa
 Different from the one in (\ref{completeNC}),  the first factor in the above expression could be negative, as now $\overline{\alpha}<0$. This makes discussion  on the instability of the black hole slightly more complicated than the case  $\overline{\alpha}>0$. It is convenient to decompose the above expression into two parts
  \beq
 f_1(L, \overline{\alpha})=L^4+2L^2(L^2-1)\overline{\alpha},
 \eeq
\beqa
f_2(L, \overline{\alpha},\overline{Q}, \ell )=&&2\Big(3+\overline{Q}^2(2\ell-1))+2\ell(\overline{\alpha}-1)-\overline{\alpha}\Big)\overline{\alpha},
\nonumber\\
&&-L^2\Big(1+\overline{Q}^2(1-4\overline{\alpha}+6\ell\overline{\alpha})+\overline{\alpha}(5-4\overline{\alpha}+\ell(6\overline{\alpha}-2)\overline{\alpha}\Big)\nonumber\\
&&+L^4(\ell-1)\Big(1+\overline{\alpha}+2\overline{\alpha}^2+\overline{Q}^2(2\overline{\alpha}-1)\Big),
\eeqa
then the necessary condition (\ref{minusalphaNC}) is equivalent to
\beq
f_1 \,\cdot f_2<0.
\eeq
In other words, the necessary condition for the black hole to develop the unstable mode is either
\beq
f_1>0, \hspace{3ex}\mbox{and}\hspace{3ex}f_2<0,
\eeq
or
\beq
f_1<0, \hspace{3ex}\mbox{and}\hspace{3ex}f_2>0.
\eeq
In the former case,  the discussion of $f_2<0$ is similar to the one for the case $\overline{\alpha}>0$.
\begin{figure}[t]
 \begin{center}
  \includegraphics[width=45mm,angle=0]{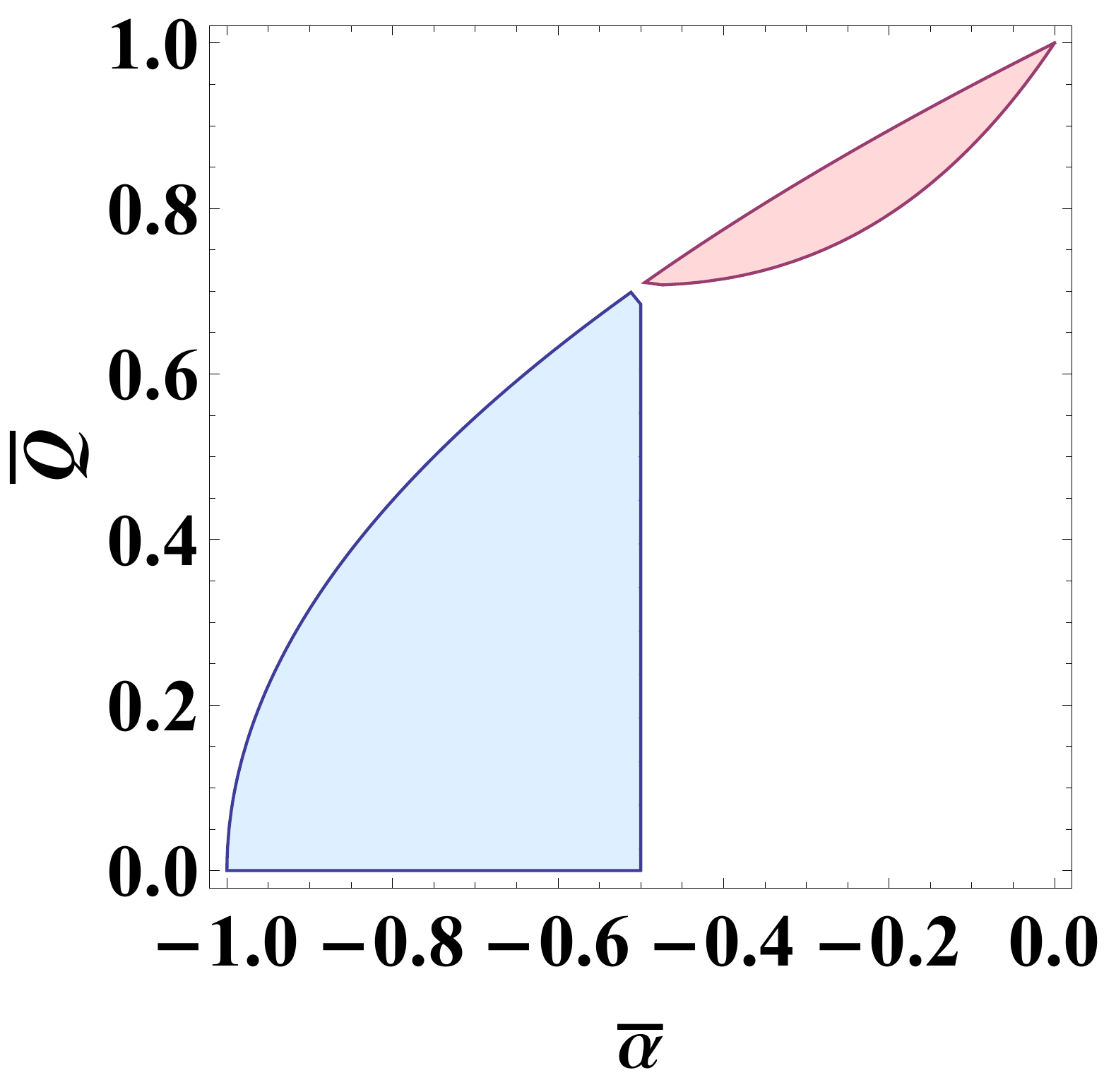}
 \end{center}
 \vspace{-5mm}
 \caption{The unstable regions of the asymptotically flat charged Gauss-Bonnet black hole are shown in $(\overline{\alpha}, \overline{Q})$ plane. The light red region corresponds to $f_2<0$ and the light blue region  corresponds to $f_1<0$, since these two regions have no overlap, both of them belong to the unstable regions.}
 \label{fig7}
\end{figure}
 \paragraph{Asymptotically flat Gauss-Bonnet black hole} For the asymptotically flat charged Gauss-Bonnet black hole, the necessary condition  (\ref{minusalphaNC}) is reduced to
\beq
(1+2\overline{\alpha})\Big(1+\overline{\alpha}+2\overline{\alpha}^2+\overline{Q}^2(2\overline{\alpha}-1)\Big)<0,
\eeq
from which it is straightforward to read the unstable regions, as shown in Fig. \ref{fig7}, where we have taken into account that $\overline{Q}^2\leq1+\overline{\alpha}$. In particular, if $\overline{\alpha}<-1/2$, the first factor $1+2\overline{\alpha}<0$ while the factor in the second bracket is positive definite such that  the black hole is always unstable, even when  $\overline{Q}=0$. On the other hand, when $\overline{\alpha}\to 0$, the unstable region becomes more and more smaller. When $\overline{\alpha}= 0$ the unstable region becomes vanishing and the black hole becomes stable, which is in accord with the fact that the asymptotical flat charged black hole in the Einstein gravity is always stable.

\begin{figure}[t]
 \begin{center}
  \includegraphics[width=35mm,angle=0]{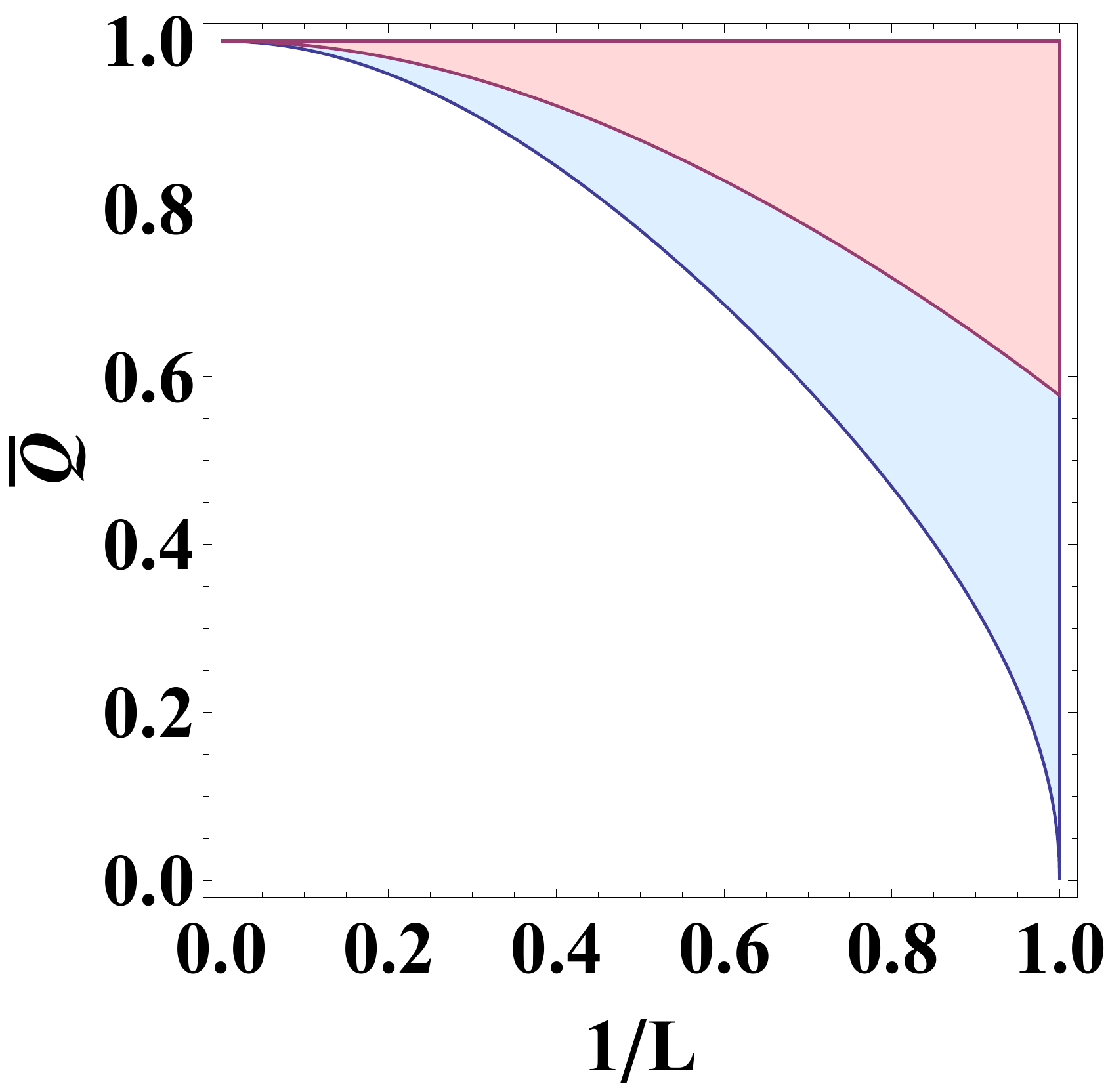}
 \hspace{1mm}
  \includegraphics[width=35mm,angle=0]{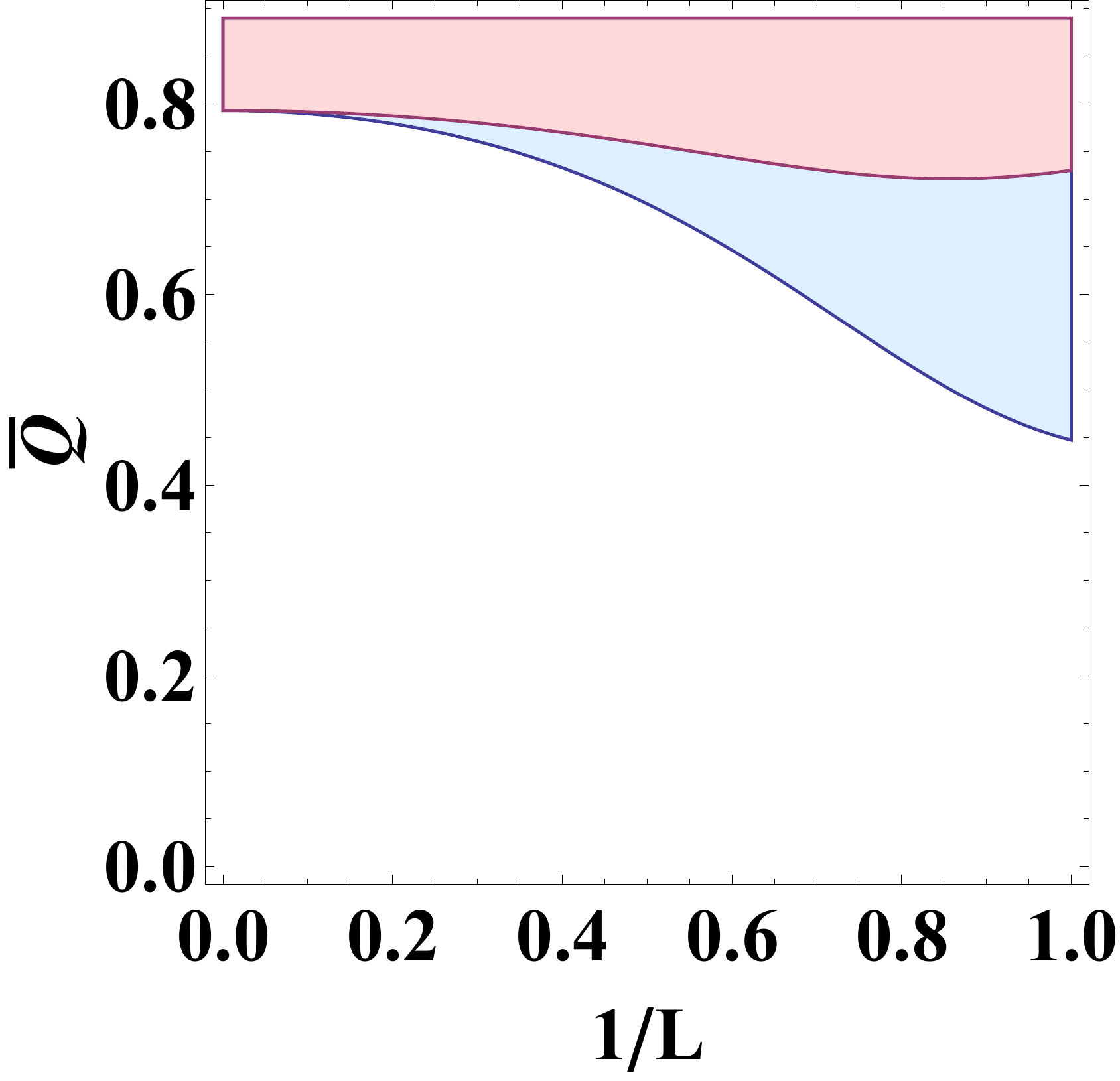}
   \hspace{1mm}
  \includegraphics[width=35mm,angle=0]{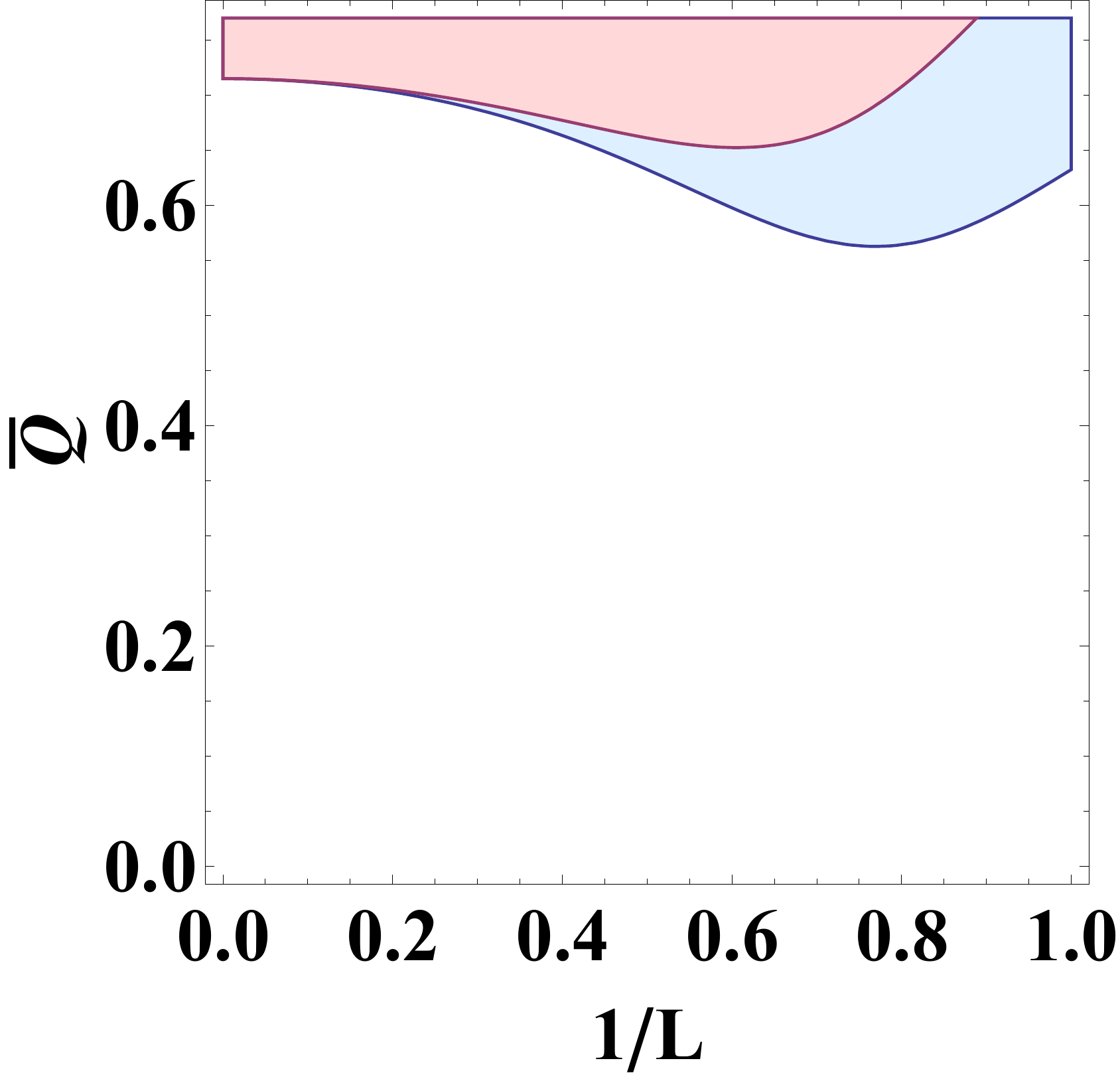}
  \hspace{1mm}
  \includegraphics[width=35mm,angle=0]{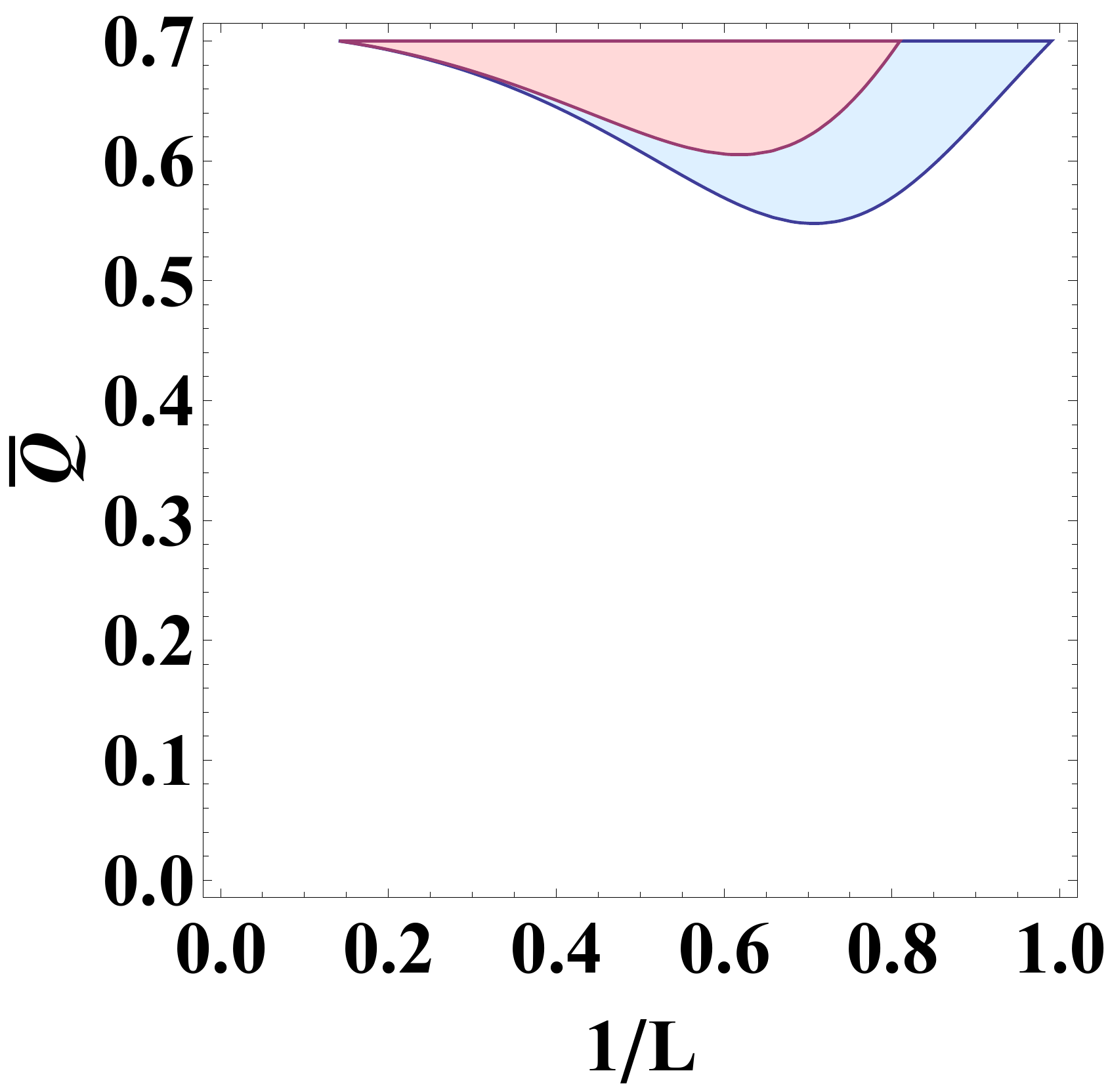}
 \end{center}
 \vspace{-5mm}
 \caption{The unstable regions of the de Sitter charged Gauss-Bonnet black hole for $\overline{\alpha}\geq-1/2$. From the left to the right $\overline{\alpha}= 0, -0.2, -0.4$ and $-0.5$, respectively, where the light red region corresponds to $\ell=3$ and the light red region plus the light blue region correspond to $\ell=2$.}
 \label{fig8}
\end{figure}

 \paragraph{de Sitter Gauss-Bonnet black hole} For the de Sitter charged Gauss-Bonnet black hole, the region $f_1<0$ is given by
\beq
-1<\overline{\alpha}<-\frac{L^2}{2(L^2-1)}.
\eeq
 Since
\beq
-\frac{L^2}{2(L^2-1)}\Bigg|_\textrm{max}=-\frac{1}{2},
\eeq
if $\overline{\alpha}\geq-1/2$, then $f_1\geq0$ is always satisfied and the unstable region is determined by $f_2<0$. In Fig. \ref{fig8} we show the unstable regions of the de Sitter charged Gauss-Bonnet
black hole with different values of  $\overline{\alpha}\geq-1/2$. Similar to the case $\overline{\alpha}>0$,   the larger angular quantum number $\ell$ reduces the
unstable region as well, and the instability occurs only when the charge is larger than a critical value.  But  now as the  Gauss-Bonnet coefficient $\overline{\alpha}$ is negative, the range of $\overline{Q}$ shrinks due to the relation (\ref{Qbar}), as one can see from Fig. \ref{fig8}.

\begin{figure}[t]
 \begin{center}
  \includegraphics[width=35mm,angle=0]{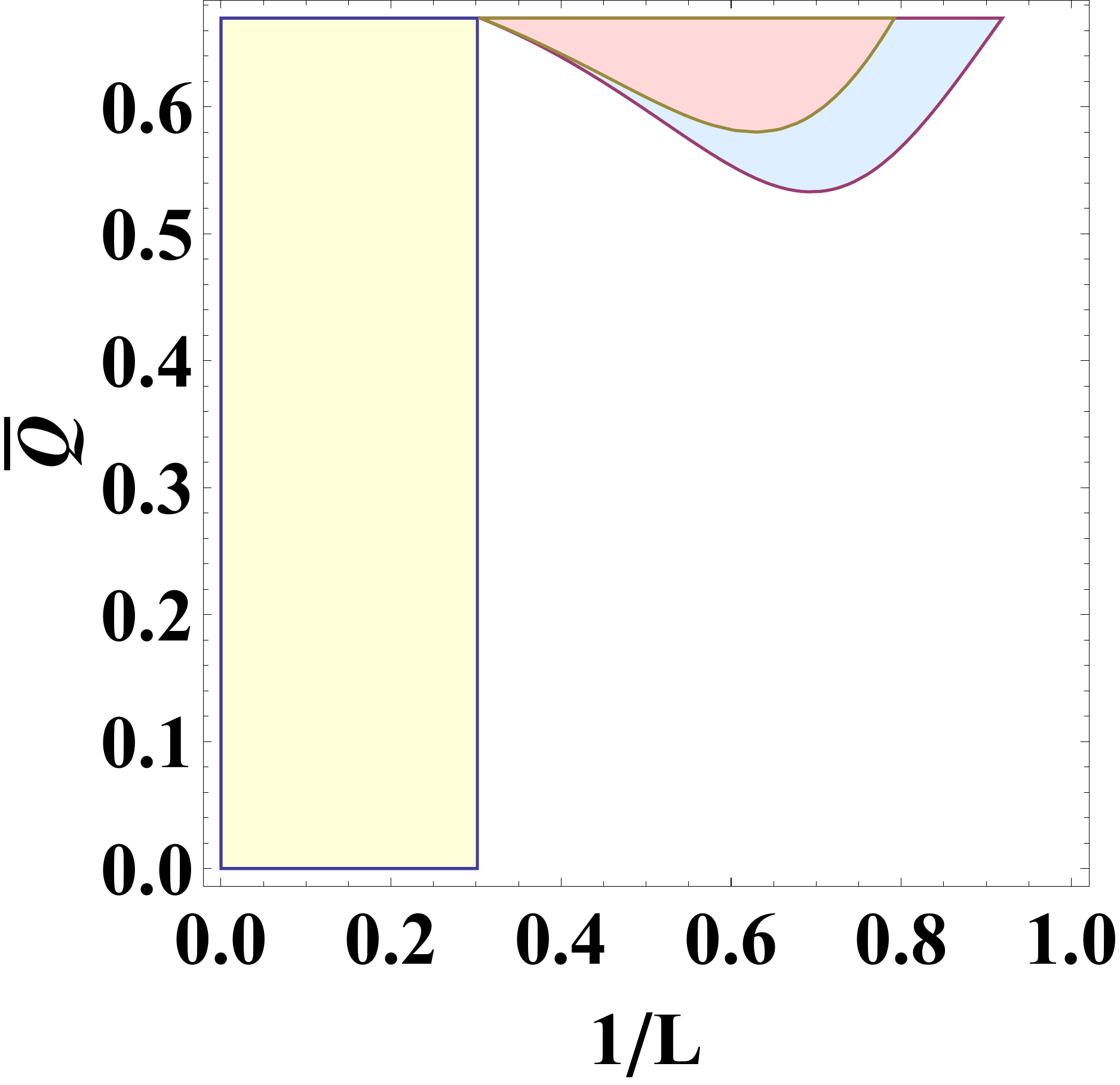}
 \hspace{1mm}
  \includegraphics[width=35mm,angle=0]{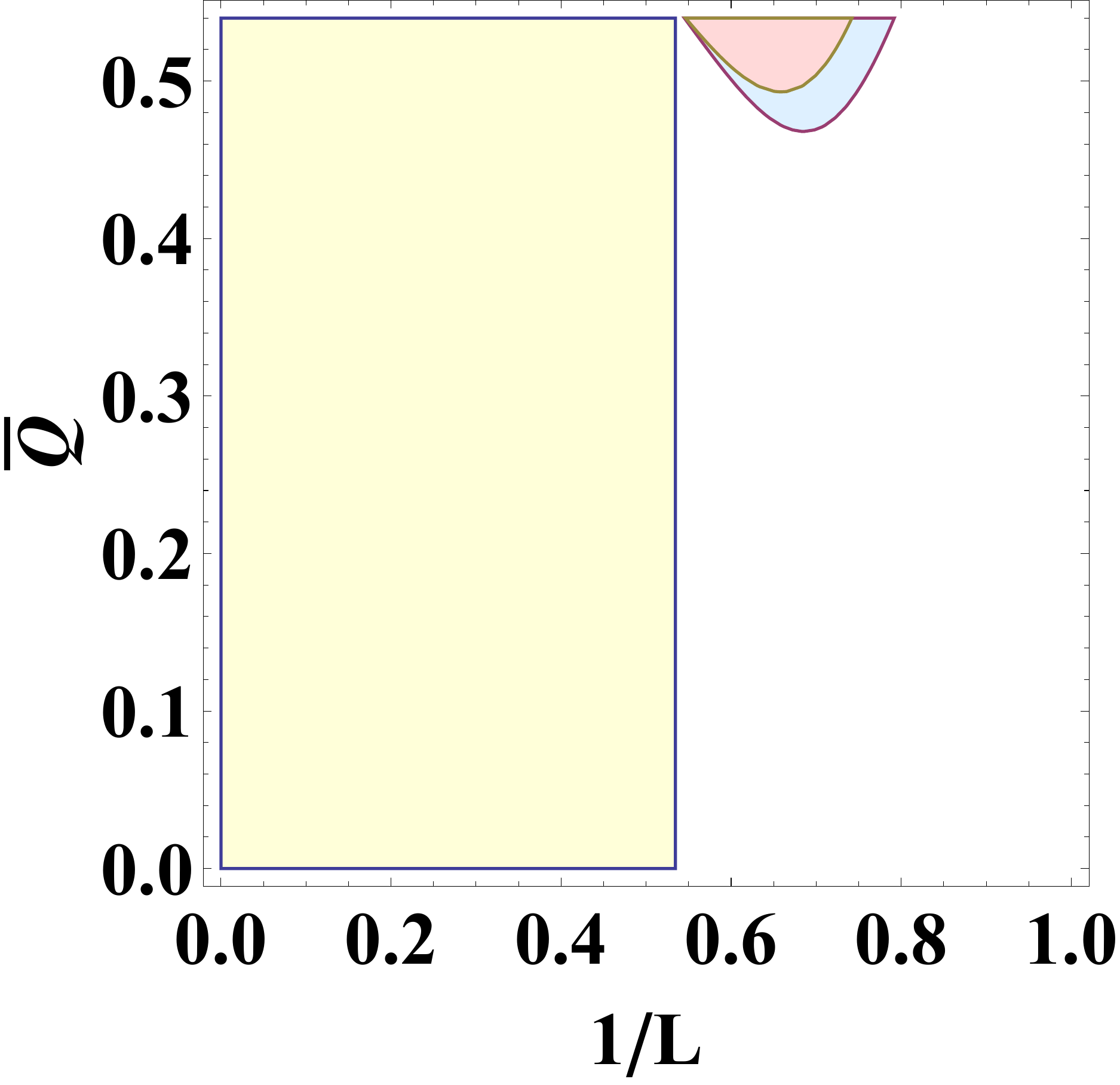}
   \hspace{1mm}
  \includegraphics[width=35mm,angle=0]{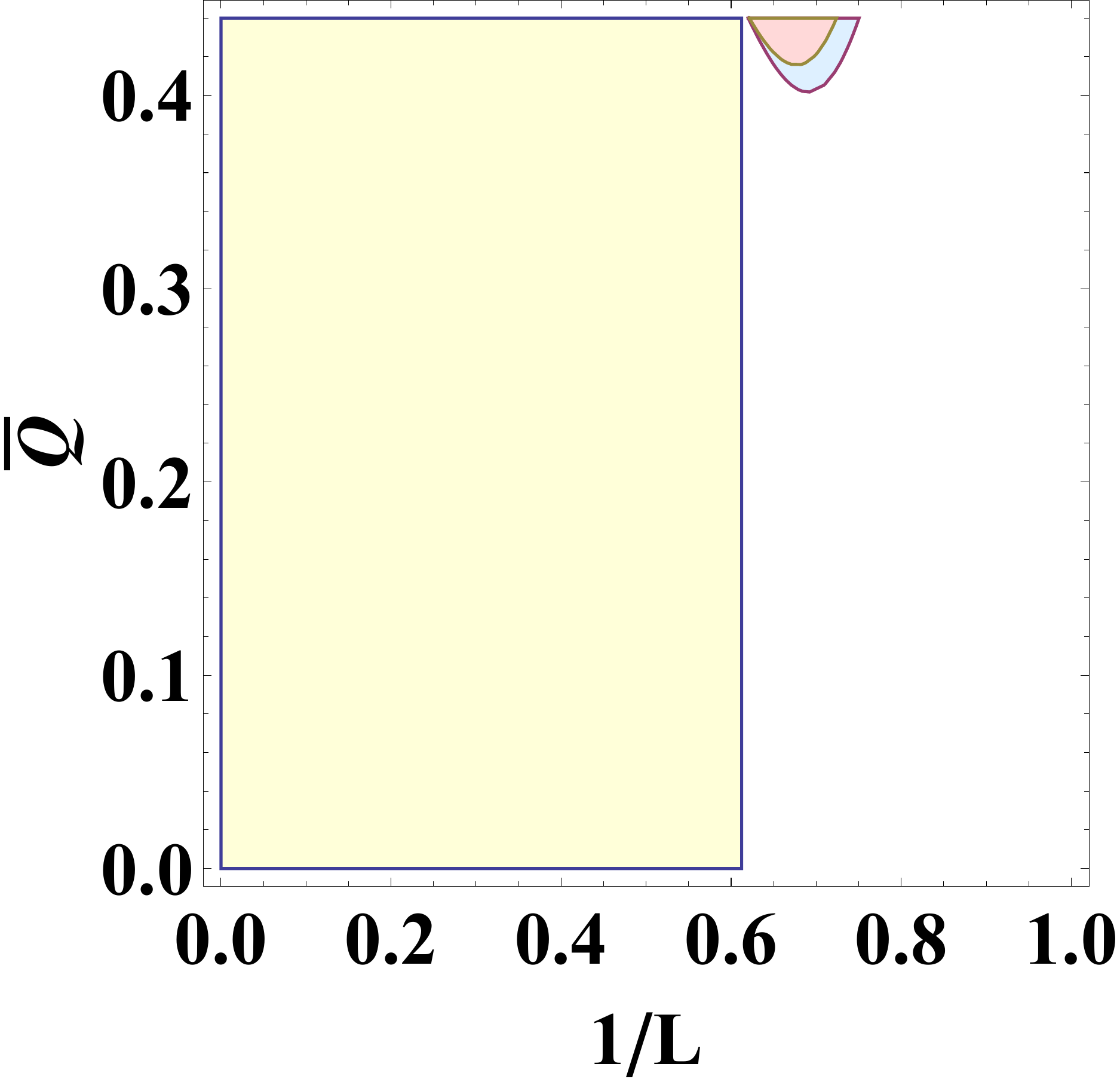}
  \hspace{1mm}
  \includegraphics[width=35mm,angle=0]{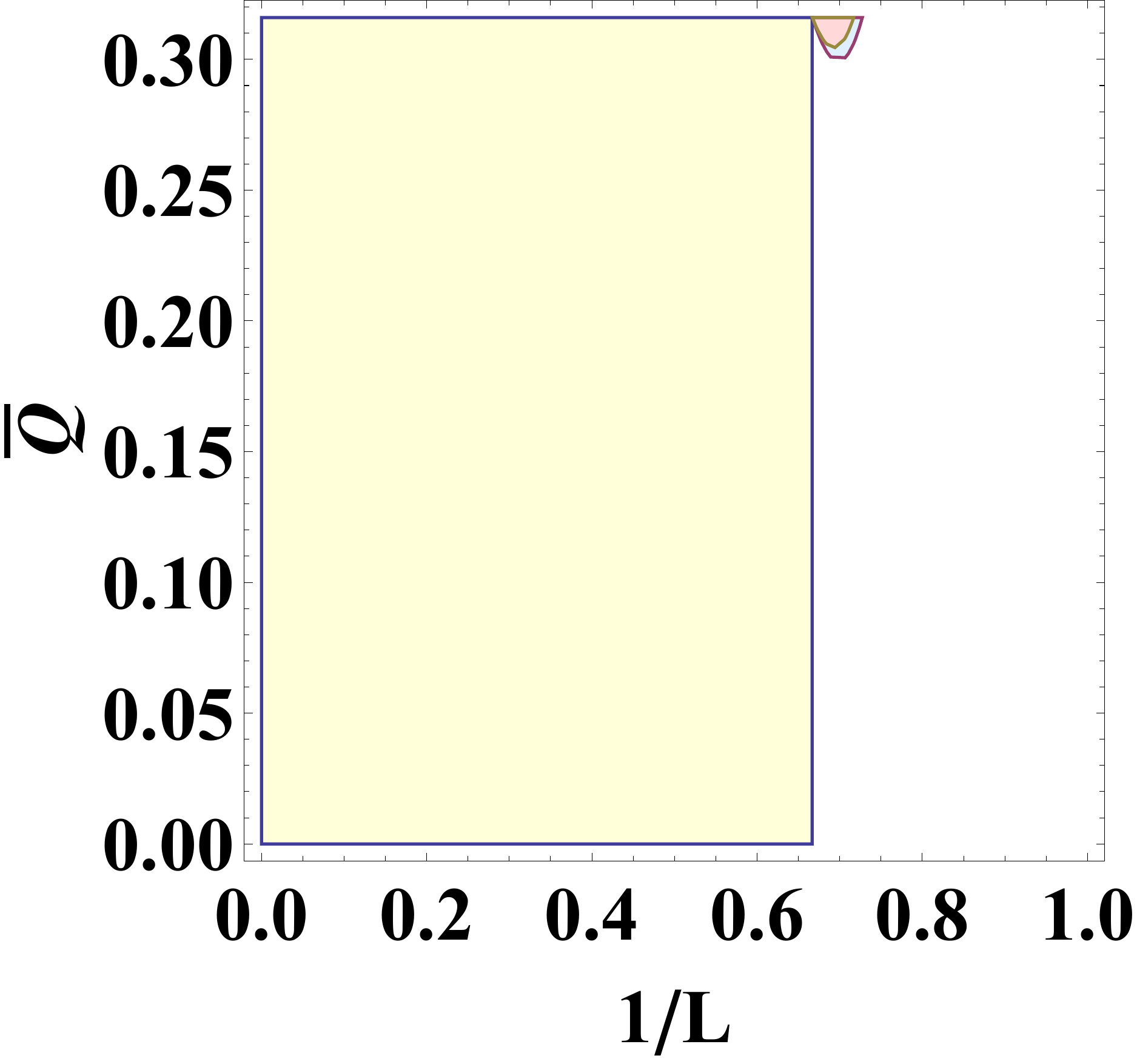}
 \end{center}
 \vspace{-5mm}
 \caption{The unstable regions of the de Sitter charged Gauss-Bonnet black hole for $\overline{\alpha}<-1/2$. From the left to the right $\overline{\alpha}= -0.55, -0.7, -0.8$ and $-0.9$, respectively, where the light yellow region corresponds to $f_1<0$ and the light red region plus the light blue region correspond to $f_2<0$. As in Fig. \ref{fig8}, the light red region corresponds to $\ell=3$ and the light red region plus the light blue region correspond to $\ell=2$.}
 \label{fig9}
\end{figure}
If  $\overline{\alpha}<-1/2$, then  $f_1$ could be negative. When $f_1$ is negative, as
 \beq
 f_2\Bigg|_{\frac{1}{L^2}=1+\frac{1}{2\overline{\alpha}}}=\frac{4(\ell-1)(\overline{Q}^2-1-\overline{\alpha})\overline{\alpha}}{(1+2\overline{\alpha})^2}\geq0,
 \eeq
so $f_1<0$ has no overlap with $f_2<0$, both of them belonging to the unstable regions, as shown in Fig. \ref{fig9}. For the unstable region  $f_2<0$,
 the angular quantum number $\ell$ has the same effect on the instability as the one in the case $\overline{\alpha}\geq-1/2$: the larger $\ell$, the smaller the unstable region. On the other hand, the unstable region $f_1<0$  is determined by the cosmological constant $1/L^2$: the smaller $1/L^2$, the smaller  the unstable region.

\begin{figure}[t]
 \begin{center}
  \includegraphics[width=35mm,angle=0]{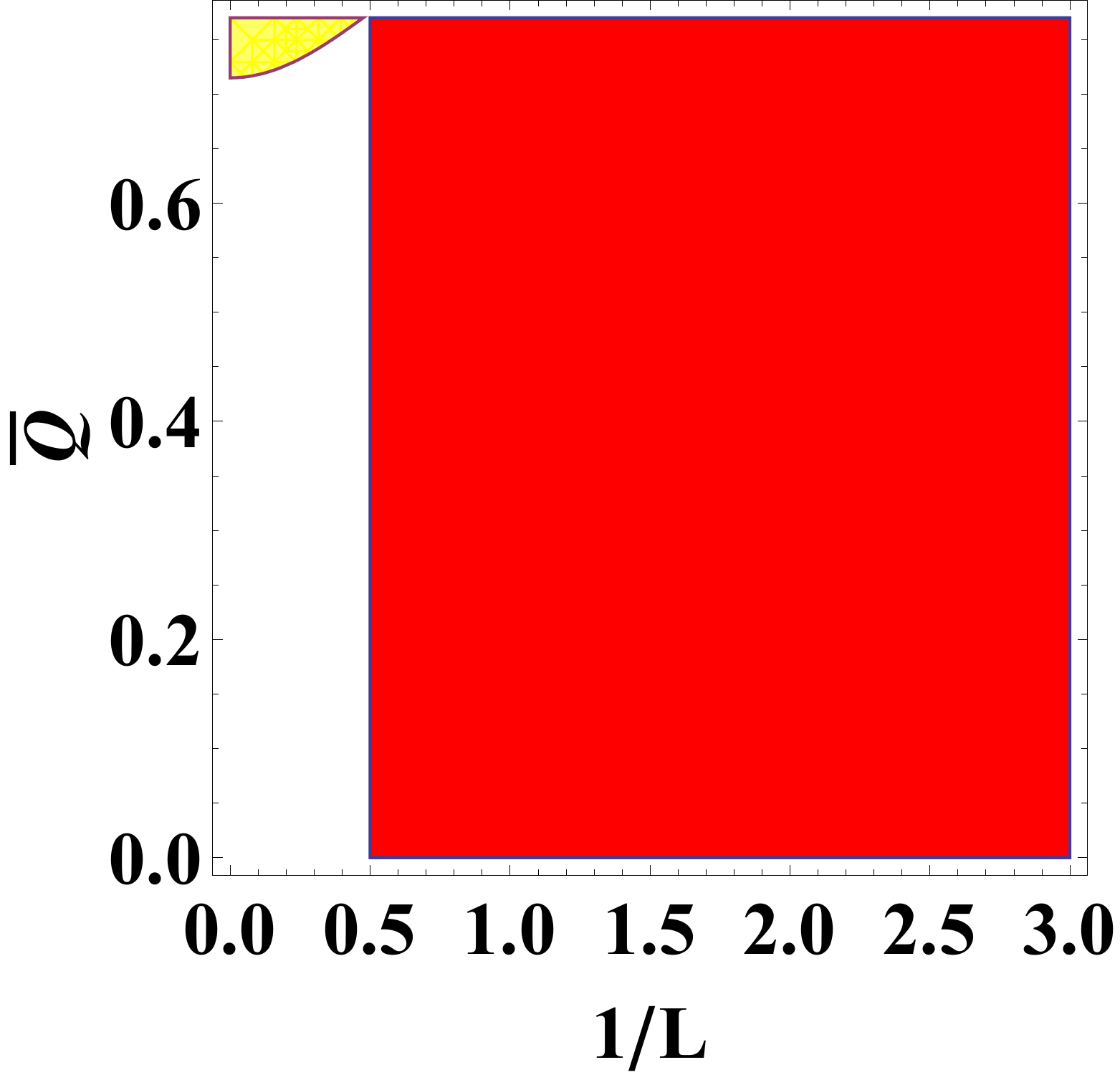}
 \hspace{1mm}
  \includegraphics[width=35mm,angle=0]{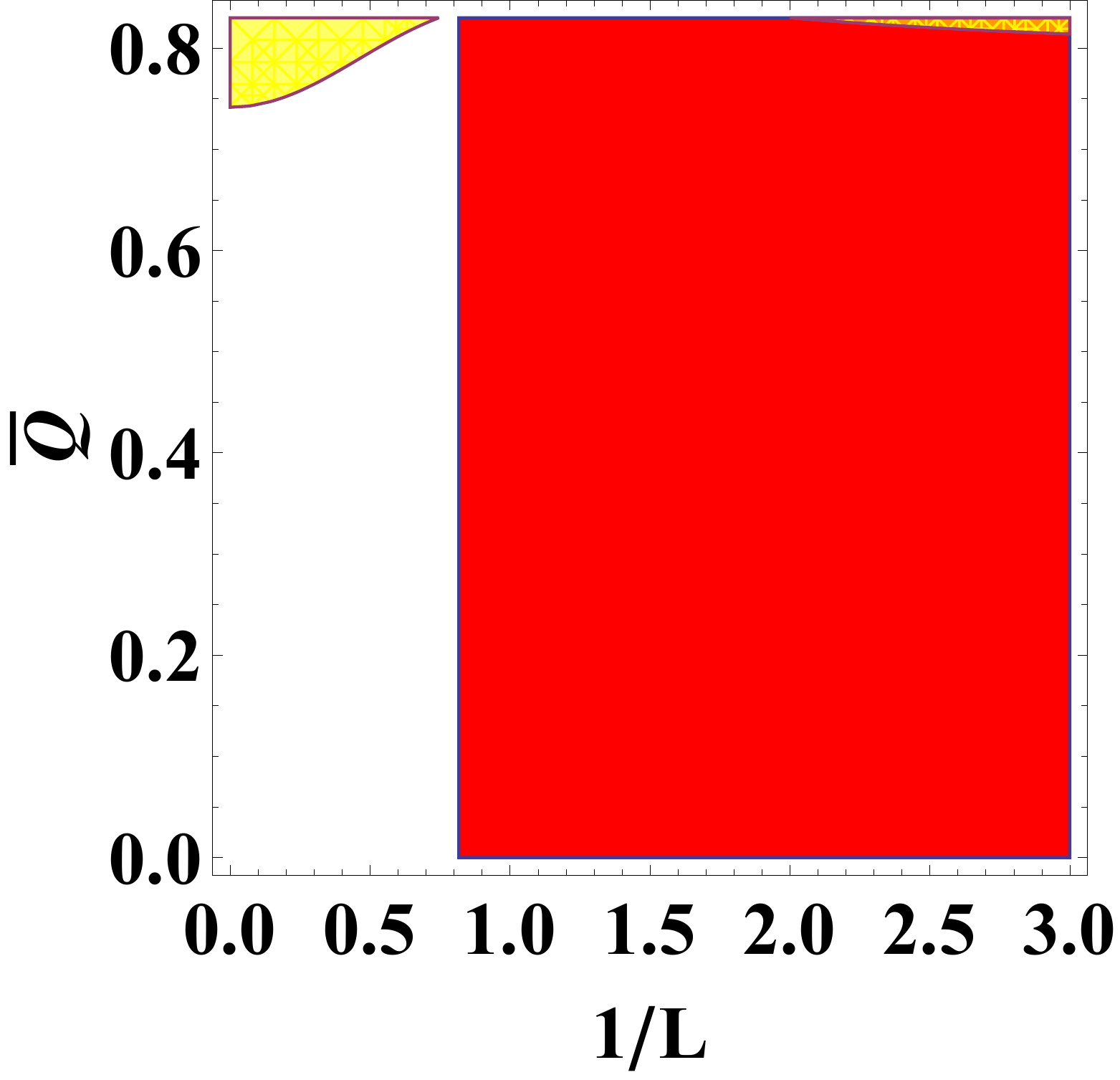}
   \hspace{1mm}
  \includegraphics[width=35mm,angle=0]{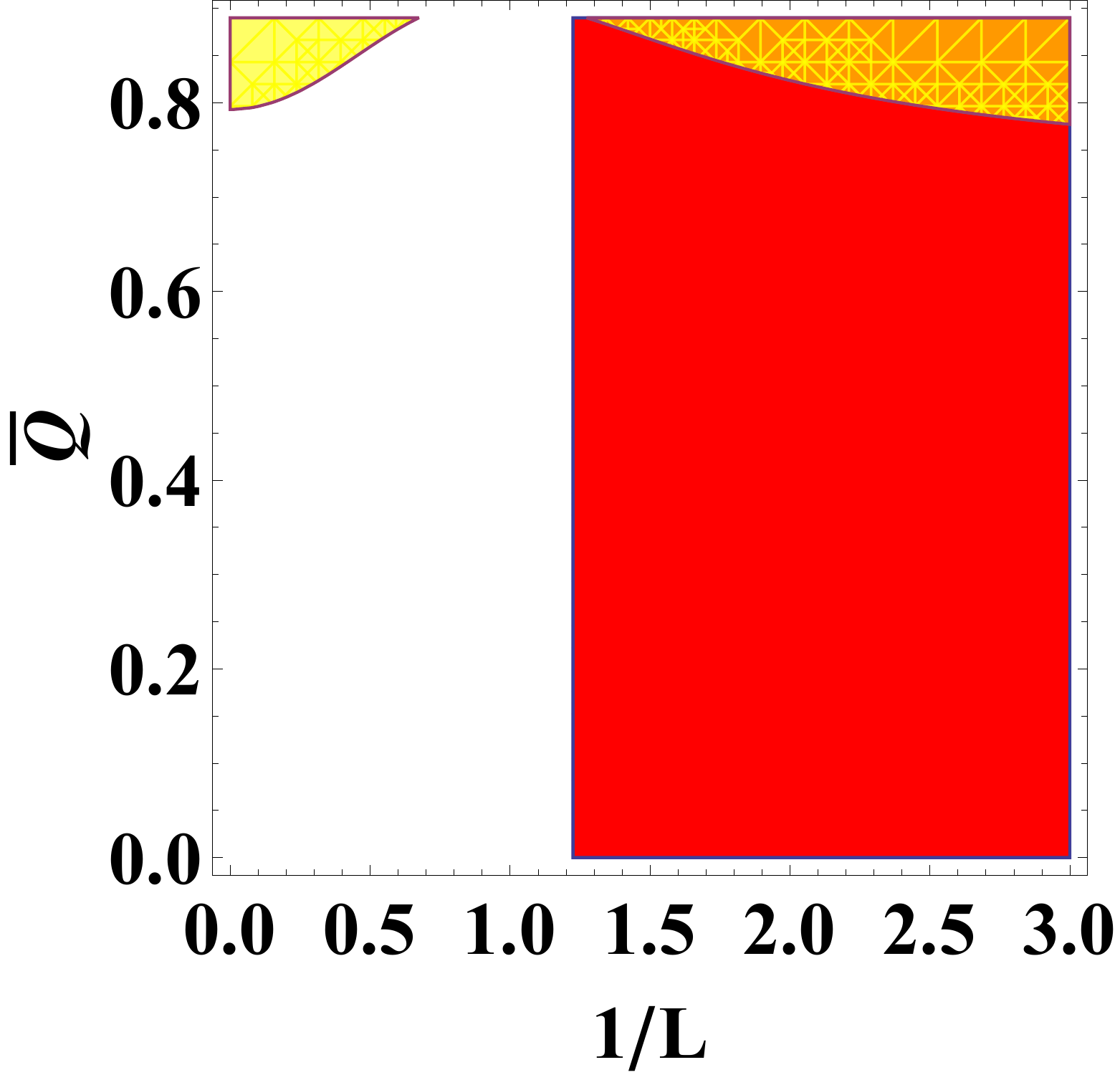}
  \hspace{1mm}
  \includegraphics[width=35mm,angle=0]{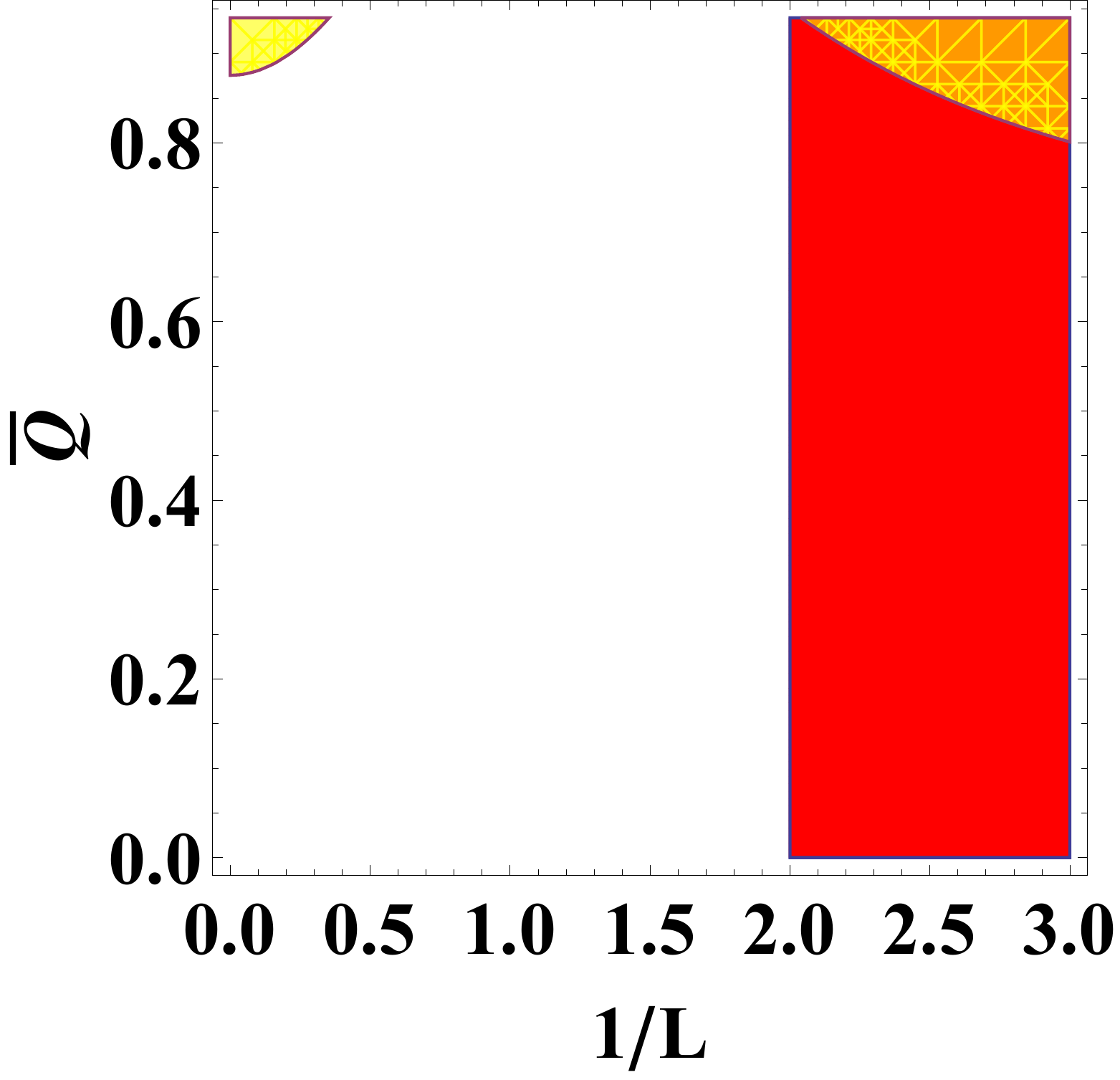}
 \end{center}
 \vspace{-5mm}
 \caption{The unstable regions of the AdS charged Gauss-Bonnet black hole. From the left to the right $\overline{\alpha}= -0.4, -0.3, -0.2$ and $-0.1$, respectively, where the red region corresponds to $f_1<0$, the yellow region correspond to $f_2<0$ and the orange region is the overlap which belongs to the stable region. Here we take $\ell=2$.}
 \label{fig10}
\end{figure}
\begin{figure}[t]
 \begin{center}
  \includegraphics[width=35mm,angle=0]{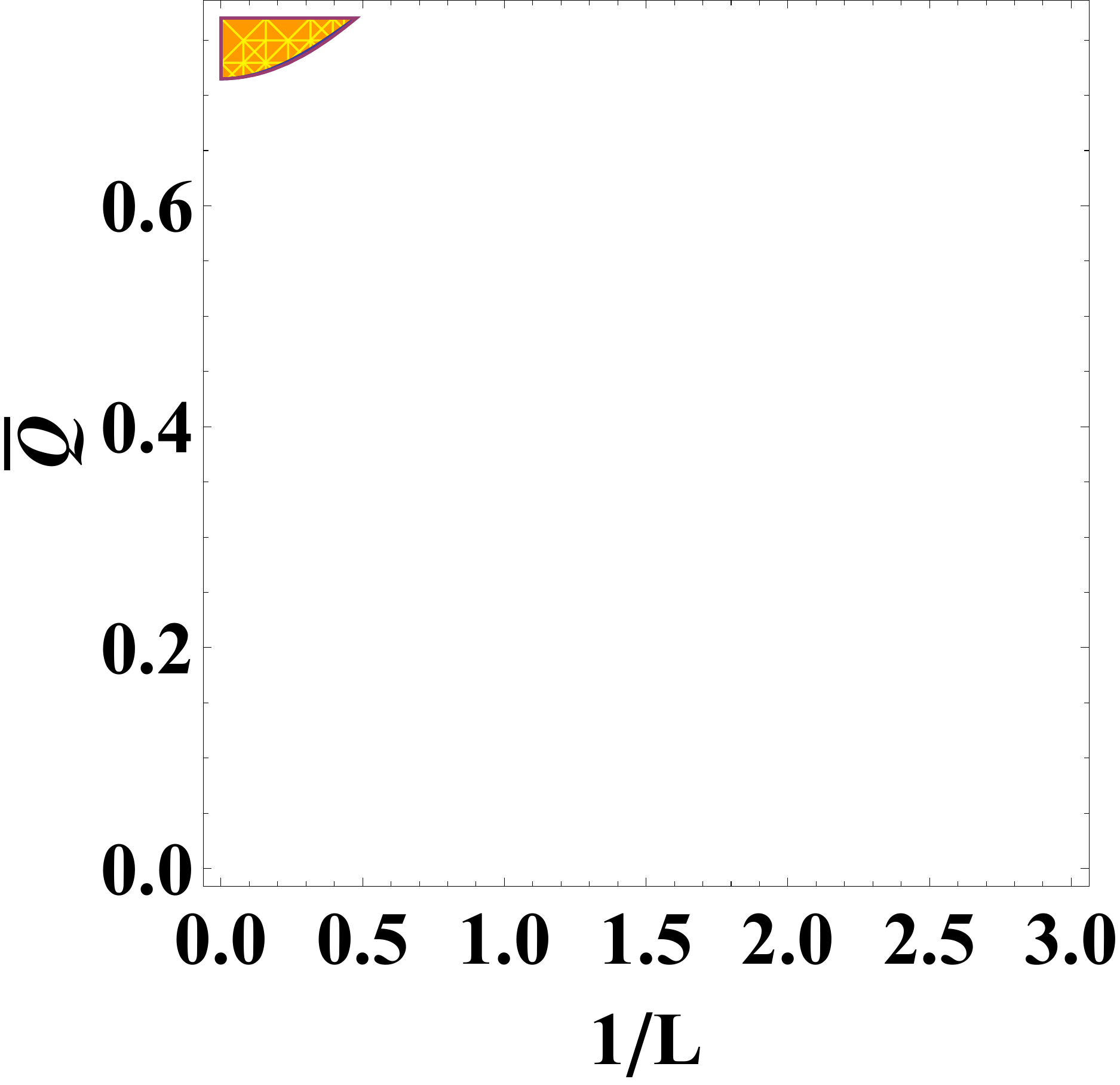}
 \hspace{1mm}
  \includegraphics[width=35mm,angle=0]{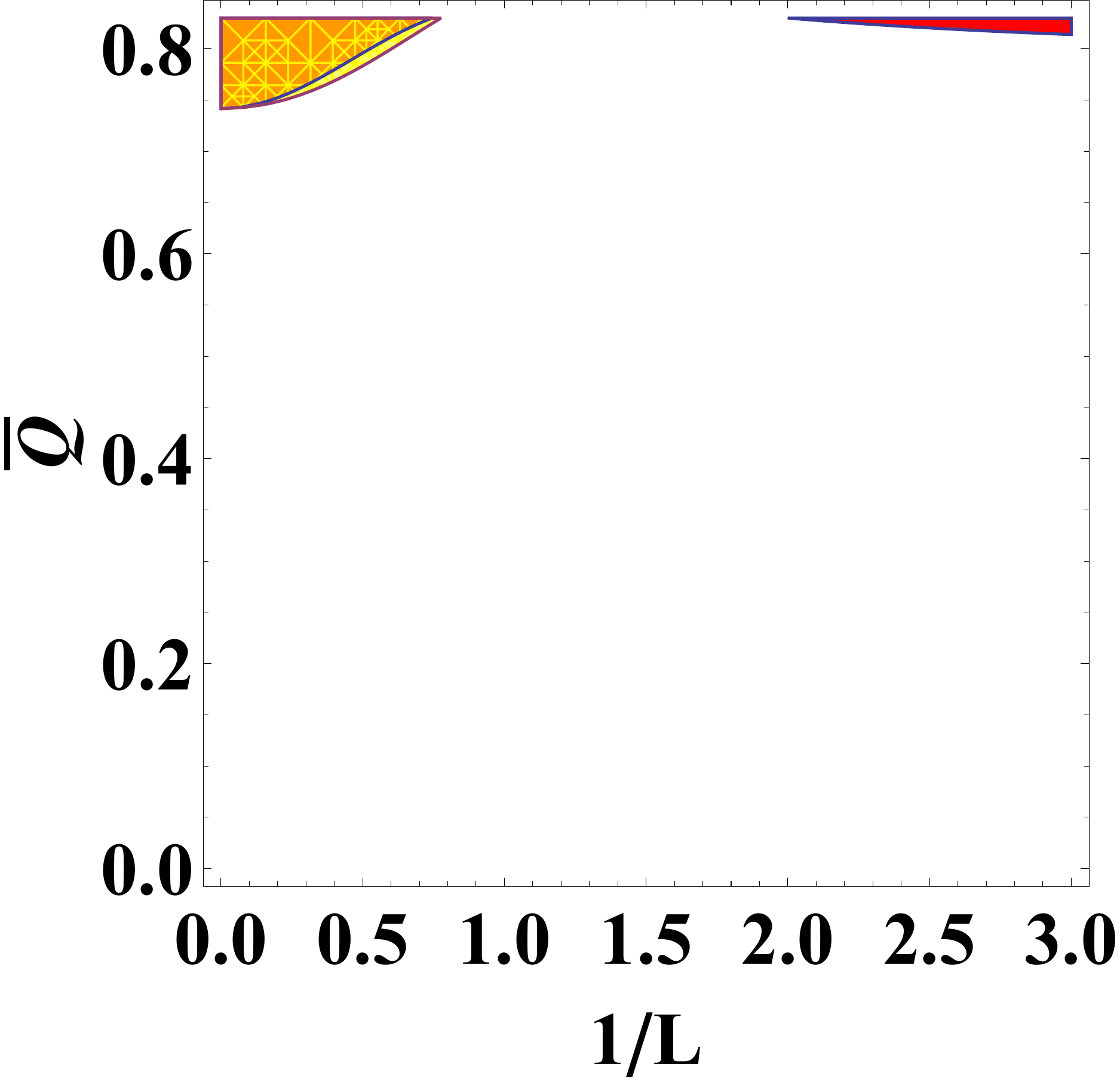}
   \hspace{1mm}
  \includegraphics[width=35mm,angle=0]{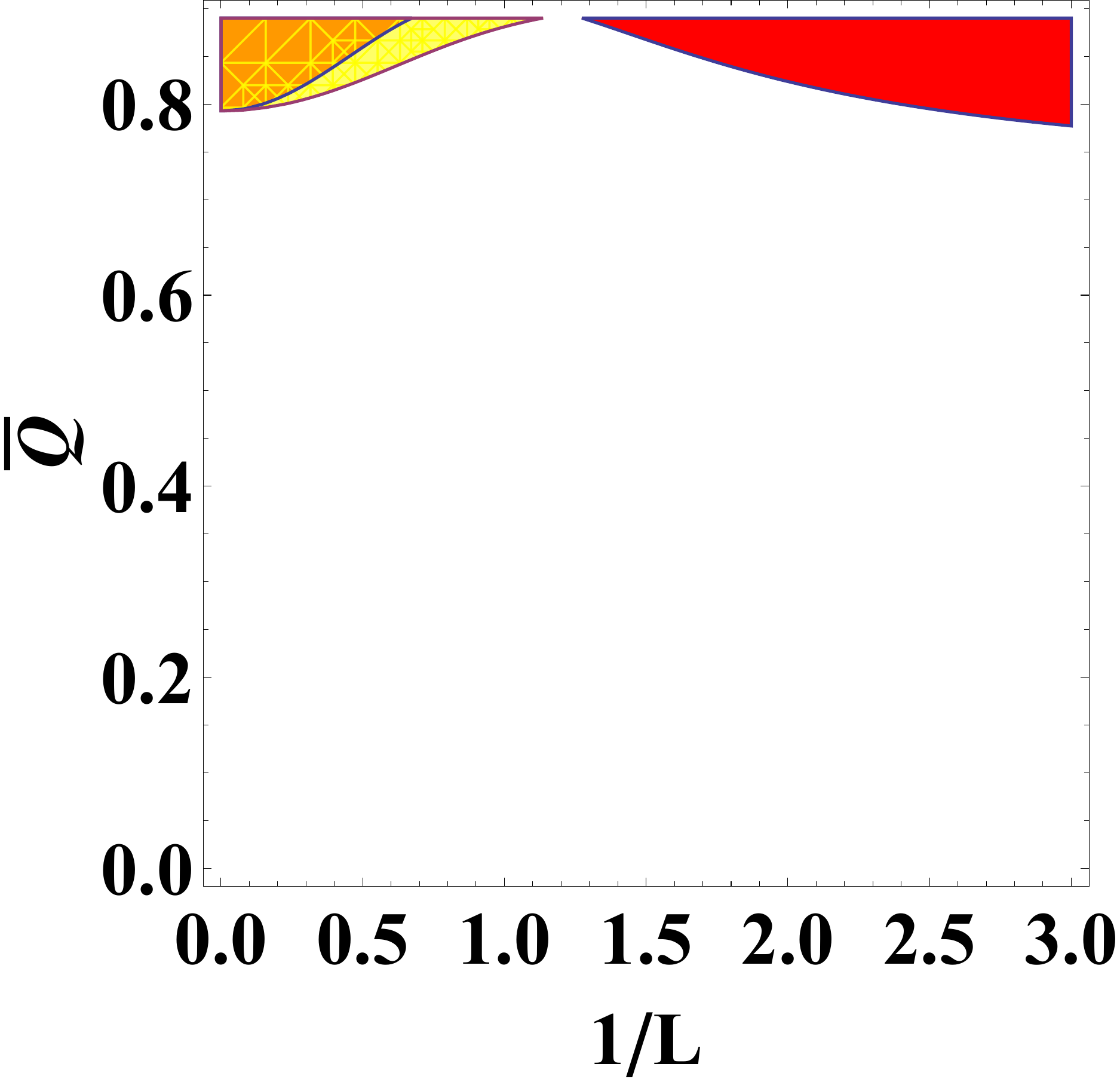}
  \hspace{1mm}
  \includegraphics[width=35mm,angle=0]{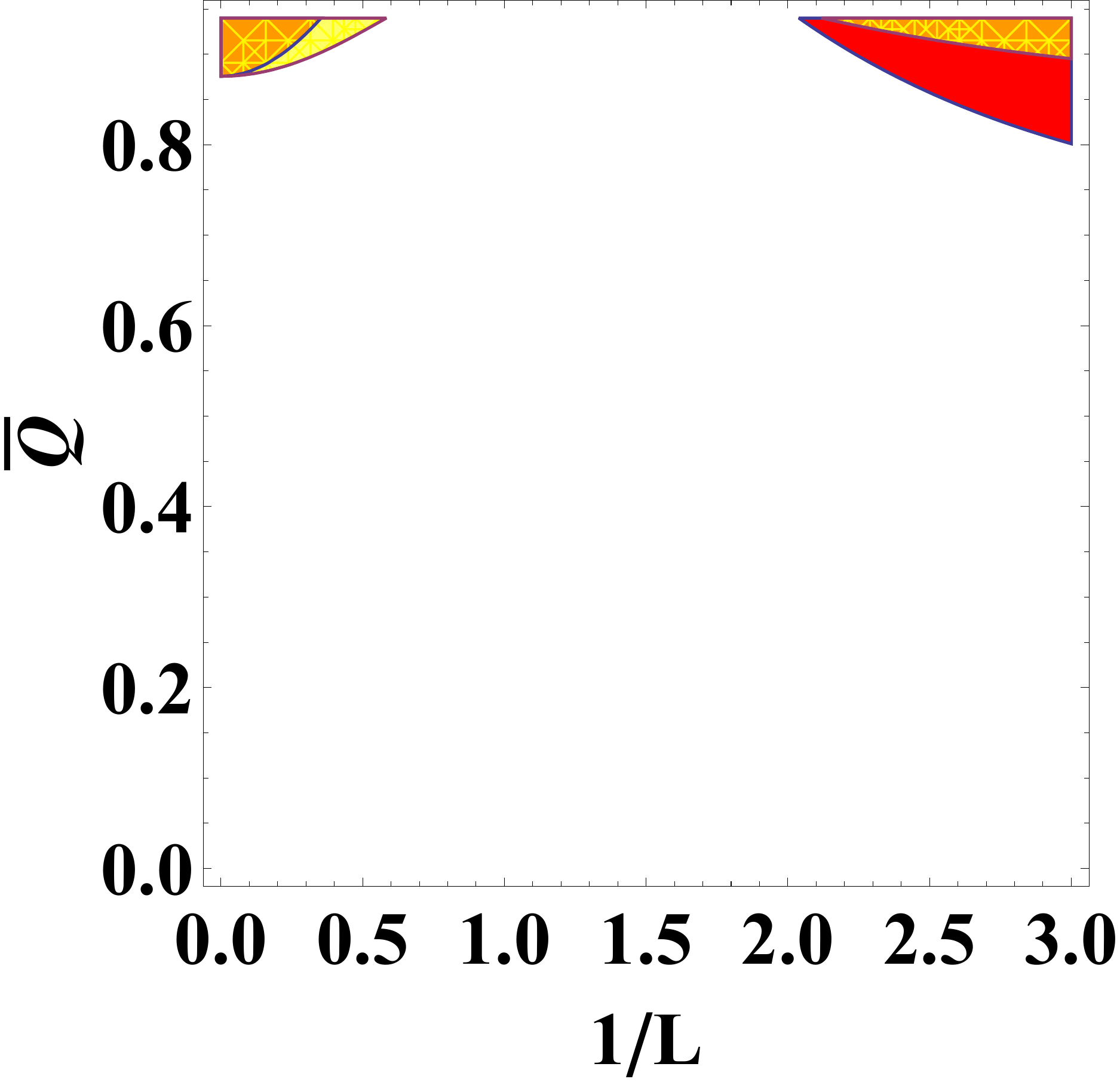}
 \end{center}
 \vspace{-5mm}
 \caption{The  region  $f_2<0$ with $\ell=2$ and $\ell=3$. From the left to the right $\overline{\alpha}= -0.4, -0.3, -0.2$ and $-0.1$, respectively, where the red region corresponds to $\ell=2$, the yellow region correspond to $\ell=3$. The upper left corner shows that a larger $\ell$ leads to a larger unstable region.  On the contrary  the right upper corner shows that a  larger $\ell$ corresponds to a smaller region. However, the right upper corner is the region overlapping  with $f_1<0$ so that it actually belongs to the stable region.  In any case, a larger $\ell$ always leads to a smaller unstable region. }
 \label{fig11}
\end{figure}
\begin{figure}[t]
 \begin{center}
  \includegraphics[width=35mm,angle=0]{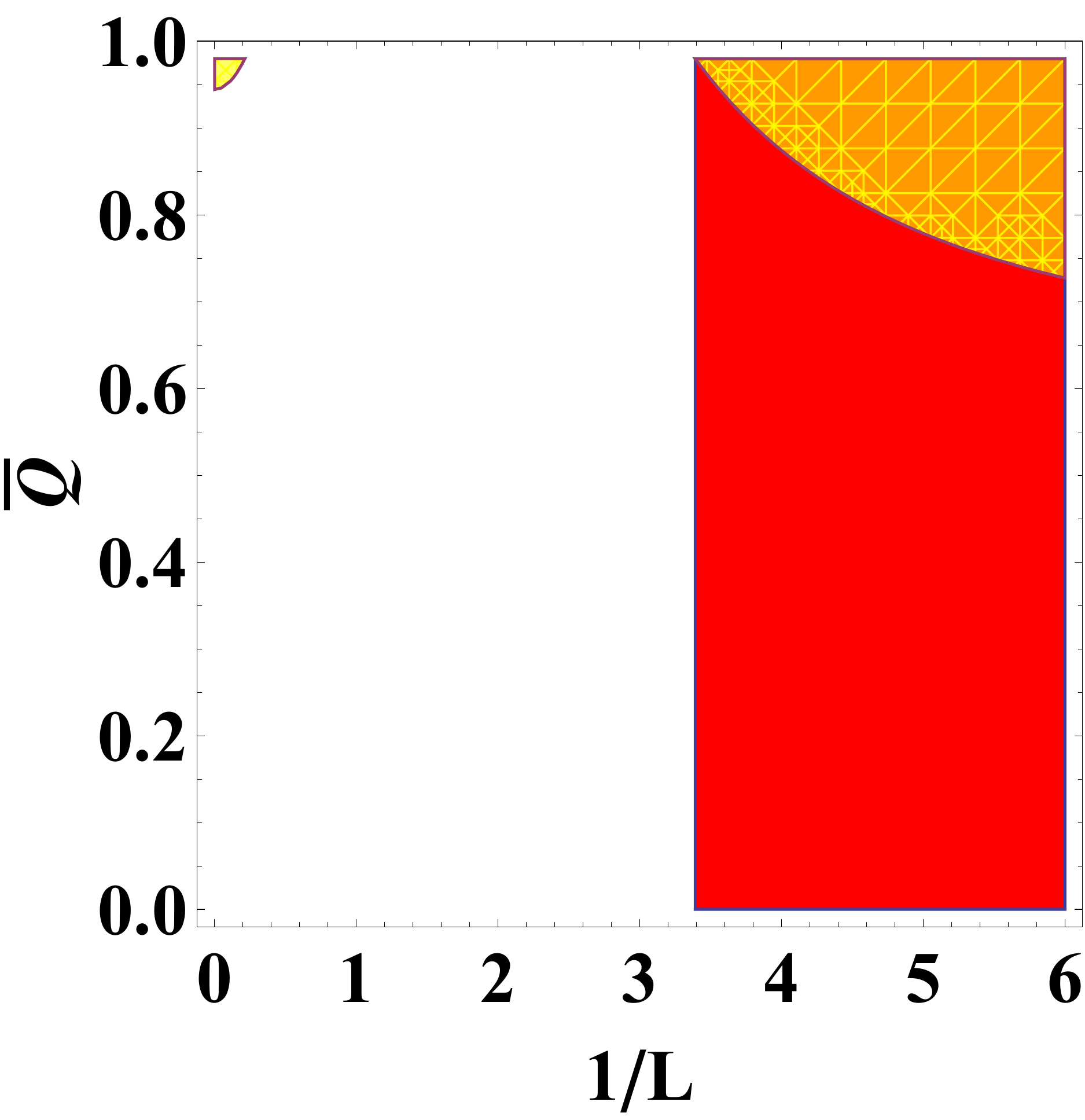}
 \hspace{1mm}
  \includegraphics[width=35mm,angle=0]{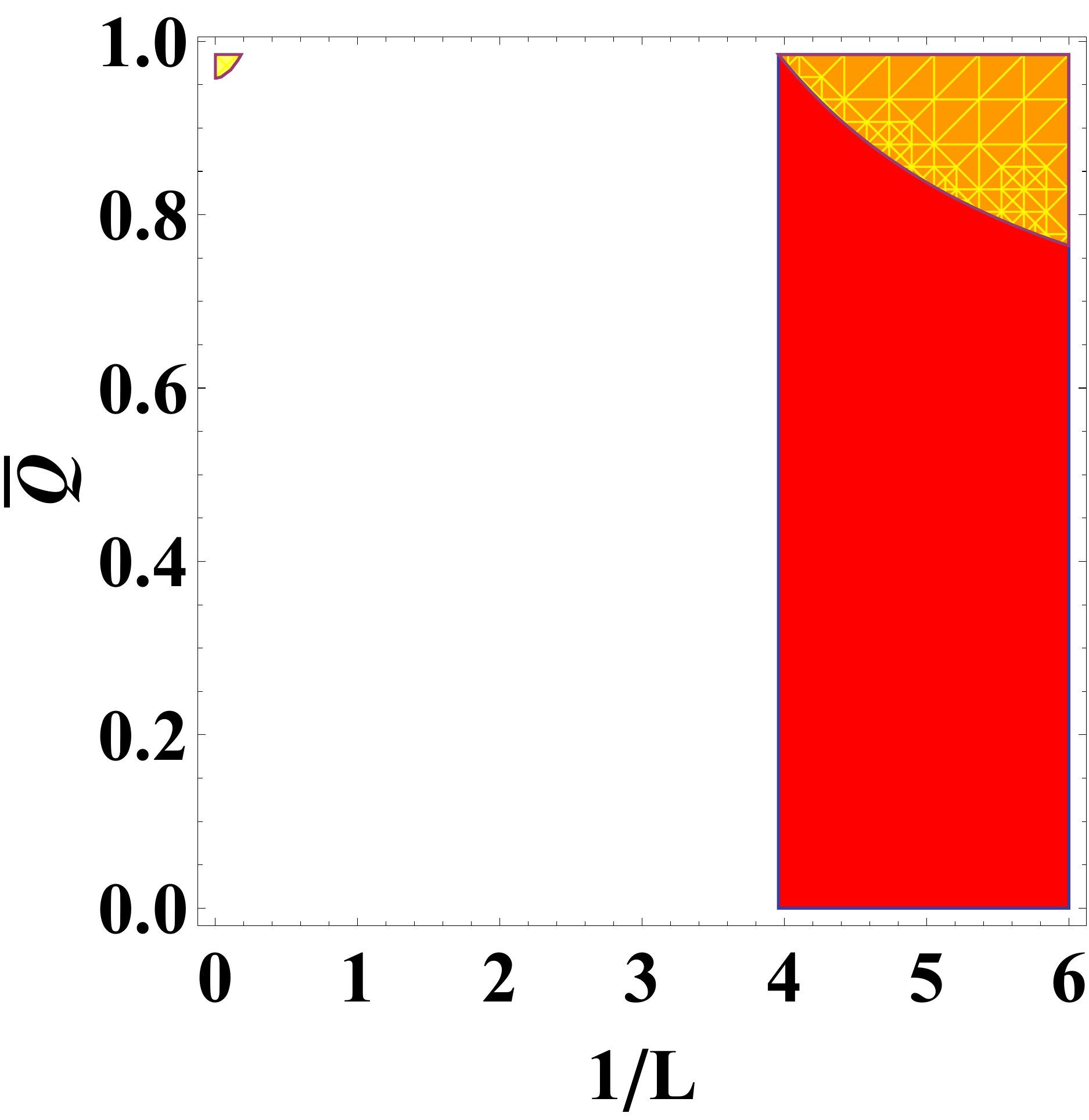}
   \hspace{1mm}
  \includegraphics[width=35mm,angle=0]{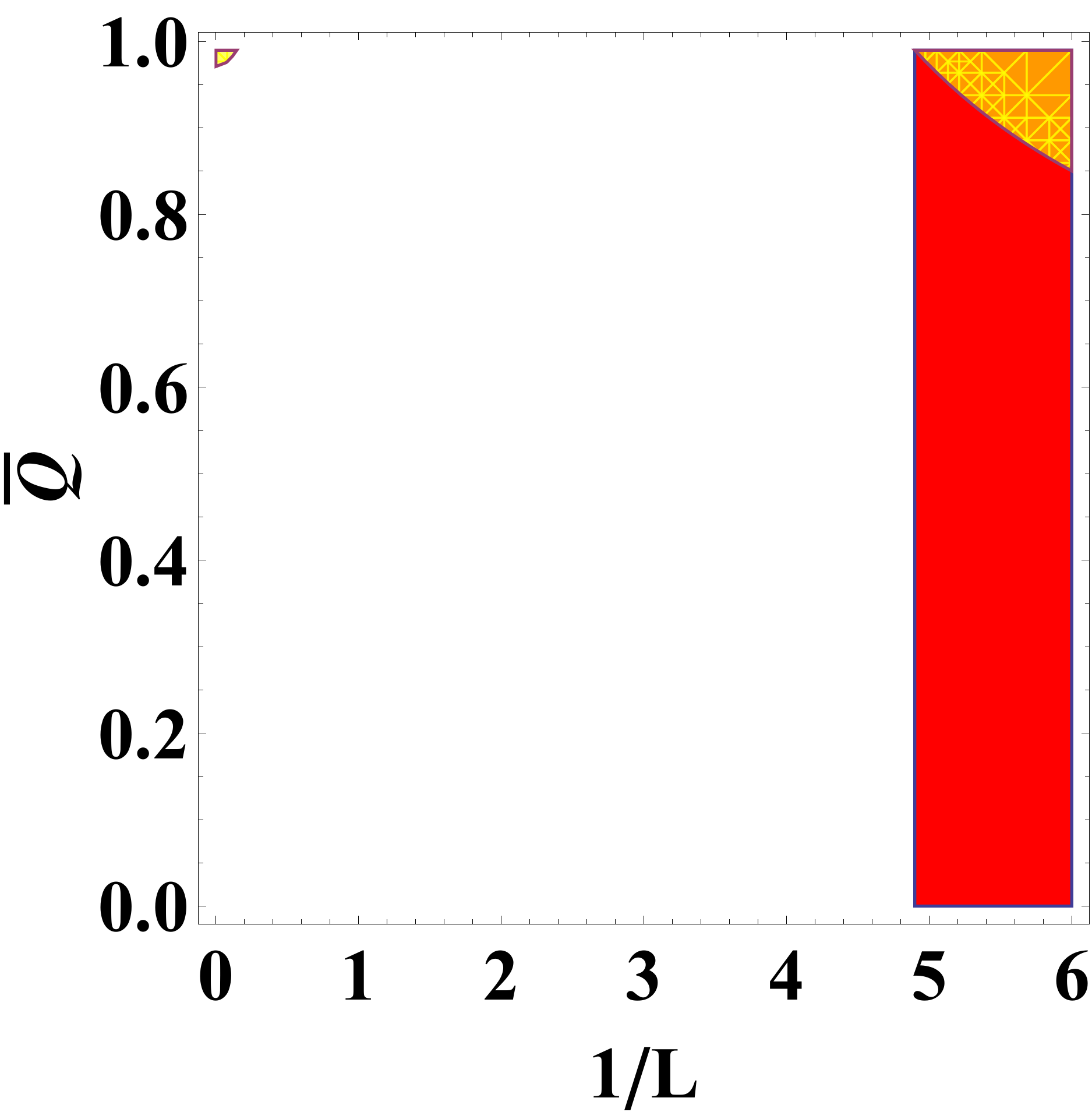}
  \hspace{1mm}
  \includegraphics[width=35mm,angle=0]{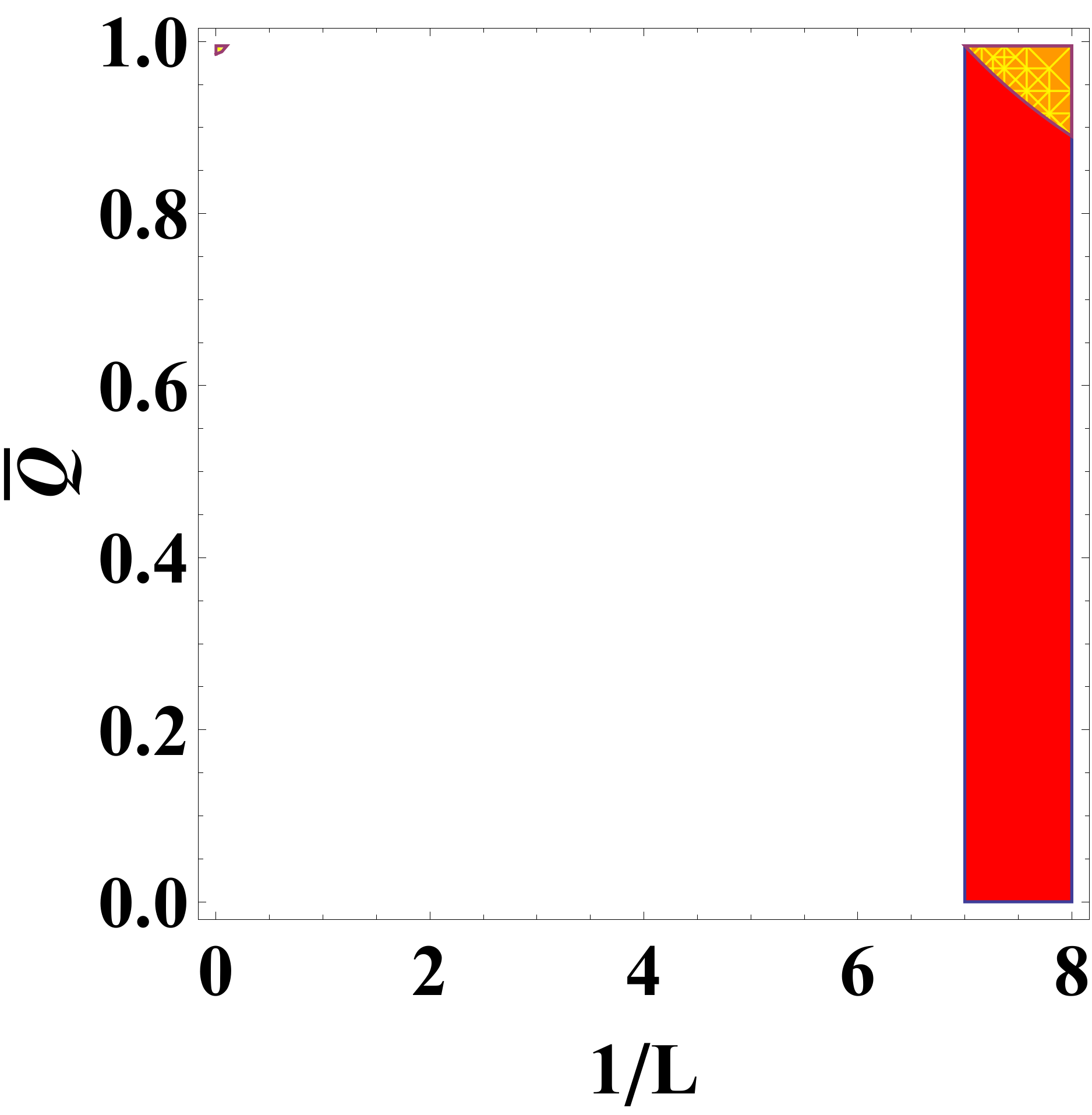}
 \end{center}
 \vspace{-5mm}
 \caption{The unstable regions of the AdS charged Gauss-Bonnet black hole. From the left to the right $\overline{\alpha}= -0.04, -0.03, -0.02$ and $-0.01$, respectively, where the red region corresponds to $f_1<0$, the yellow region correspond to $f_2<0$ and the orange region is their overlap which belongs to the stable region. Here we take $\ell=2$.}
 \label{fig12}
\end{figure}
 \paragraph{AdS Gauss-Bonnet black hole} To discuss the instability of the AdS charged Gauss-Bonnet black hole,  we just need to replace $L$ by $iL$ as before.  The necessary condition for the existence of
unstable mode is still  given by $f_1 \,\cdot f_2<0$, with $f_1$ and $f_2$ being now defined by
 \beq
 f_1(L, \overline{\alpha})=L^4+2L^2(L^2+1)\overline{\alpha},
 \eeq
\beqa\label{f2}
f_2(L, \overline{\alpha},\overline{Q}, \ell )=&&2\Big(3+\overline{Q}^2(2\ell-1))+2\ell(\overline{\alpha}-1)-\overline{\alpha}\Big)\overline{\alpha},
\nonumber\\
&&+L^2\Big(1+\overline{Q}^2(1-4\overline{\alpha}+6\ell\overline{\alpha})+\overline{\alpha}(5-4\overline{\alpha}+\ell(6\overline{\alpha}-2)\overline{\alpha}\Big)\nonumber\\
&&+L^4(\ell-1)\Big(1+\overline{\alpha}+2\overline{\alpha}^2+\overline{Q}^2(2\overline{\alpha}-1)\Big).
\eeqa
The region $f_1<0$ is given by
\beq
-1<\overline{\alpha}<-\frac{L^2}{2(L^2+1)},
\eeq
as now we only demand $L^2>0$, then
\beq
-\frac{L^2}{2(L^2+1)}\Bigg|_\textrm{min}=-\frac{1}{2}.
\eeq
If $\overline{\alpha}\leq-1/2$, then $f_1\leq0$ is always satisfied, and in this case the unstable region is completely determined by $f_2>0$.  It turns out that as long as $\overline{\alpha}\leq-1/2$, $f_2>0$
is always satisfied for the whole range of the parameters. This can be seen as follows, from (\ref{f2}) we observe that the coefficients of the linear $\overline{Q}^2$ terms are all negative, so
\beqa
f_2\geq &&f_2(Q^2=1+\overline{\alpha})\nonumber\\
&&=2\Big( L^2+2(1+L^2)\overline{\alpha}\Big)\Big(1+\big(-1+L^2(\ell-1)+2\ell\big)\overline{\alpha} \Big).
\eeqa
It is easy to determine that the minimum of the right hand side of the above expression occurs at $\overline{\alpha}=-1/2$, such that
\beq
f_2\geq f_2(Q^2=1+\overline{\alpha},\, \overline{\alpha}=-1/2)=L^2(\ell-1)+2\ell-3>0.
\eeq
If $\overline{\alpha}\geq-1/2$, different from the discussion in the de Sitter case,
 $f_1<0$ and $f_2<0$ have overlapping region when both $L$ and $|\alpha|$ are small, as shown in Fig. \ref{fig10}. Such overlapping region corresponds to stable black holes. To see the effect of the angular quantum number $\ell$ on the instability, we depict the region $f_2<0$ with different $\ell$ in Fig. \ref{fig11}, from which we can see that a larger $\ell$ leads to a larger unstable region. This is different from the case with a positive Gauss-Bonnet term, in which $\ell=2$ has the largest unstable region.

 It would be interesting to consider the $\overline{\alpha}\to 0^-$ limit. According the previous discussion, when $\overline{\alpha}\geq0$, the AdS  Reissner-Nordstrom black holes are always stable.  Actually, from Fig. \ref{fig10} we can see that as $\overline{\alpha}$ turns to zero, the unstable regions shrinks. This can be seen more clearly in Fig. \ref{fig12}, where we depict the unstable regions with $\overline{\alpha}$ being close to zero.
  From Fig. \ref{fig12} we can see that when the range of $L$ is fixed, as  $\overline{\alpha}\to 0$ the unstable region shrinks to zero. However, for a fixed tiny $\overline{\alpha}$ as long as $L$ is
sufficient small the unstable region always exists.  But we note that the validity of the $1/n$ expansion requires that $1/L$ to be finite, so  $L$ cannot be arbitrary small. Therefore the  $\overline{\alpha}\to 0$ limit is in accord with the fact that the AdS  Reissner-Nordstrom black holes are always stable.

\paragraph{Deformed static solution} The discussion of the deformed static solution in the previous section can be naturally extended to the case with a negative Gauss-Bonnet term. Since in this case the unstable mode is also of a pure imaginary frequency, so identically at the edge of the  instability there exists a non-trivial zero-mode static perturbation indicating the existence of a non-spherical symmetrical symmetric static solution branch. By solving the large $D$ effective equations in the static case, we can  find such a static solution as well. The solution
is also given by (\ref{deformedsolutionpzqm}) and (\ref{deformedsolutionPz}), where $P_2$ is given by (\ref{deformedsolutionP2}) when the black hole is in the de Sitter spacetime, and
$P_2=1$ or
\beq
P_2=\frac{2\overline{\alpha}(-3+\overline{\alpha}+\overline{Q}^2)-L^2(1+\overline{Q}^2(1-4\overline{\alpha})+(5-4\overline{\alpha})\overline{\alpha})+L^4(1+\overline{\alpha}+2\overline{\alpha}^2+\overline{Q}^2(-1+2\overline{\alpha}))}
{4\overline{\alpha}(-1+\overline{Q}^2+\overline{\alpha})+2L^2\overline{\alpha}(-1+3\overline{Q}^2+3\overline{\alpha})+L^4(1+\overline{\alpha}+2\overline{\alpha}^2+\overline{Q}^2(-1+2\overline{\alpha}))},
\eeq
 when the background spacetime is asymptotically flat or AdS correspondingly.

\section{Summary}\label{summary}

In this article we studied static Gauss-Bonnet black holes at large $D$. For generality, we considered the black holes in the Einstein-Maxwell-Gauss-Bonnet theory with a cosmological constant. The black holes include the charged black hole in the asymptotically flat and (A)dS spacetimes. After deriving  the large $D$ effective equations of the black hole to the next-leading order of $1/D$, we showed that the static black hole could still be the solution of an elastic theory. Actually we found the relation $\sqrt{-g_{vv}} \,\mathcal{K}=$ constant for the membrane embedding. Different from the Einstein gravity, the constant in the relation is not simply the surface gravity. Moreover, by considering the embedding of the membrane in the spherical coordinates, we read the static solutions and furthermore investigated their instabilities.

Considering the perturbation around the  exact solutions, we could read   the charge and scalar-type quasinormal modes  of the static Gauss-Bonnet black holes analytically. The results we got can be summarized as following. If the Gauss-Bonnet term is positive, the black holes could be stable or unstable, depending on the cosmological constant, charge and the Gauss-Bonnet coefficient.
\begin{enumerate}
\item For the static asymptotically flat Gauss-Bonnet black hole, it is always stable, no matter it is charged or not, and no matter how large  the Gauss-Bonnet parameter is. This answers the unsolved question on the stability of the Gauss-Bonnet black hole studied in \cite{CFLY} in which only the black hole with a small or a large Gauss-Bonnet parameter has shown to be stable.
\item For the static asymptotically AdS Gauss-Bonnet black hole, it is always stable, no matter it is charged or not.
\item For the static asymptotically de Sitter Gauss-Bonnet black hole, it becomes unstable when the cosmological constant is sufficiently large. This kind of $\Lambda$-instability is very similar to the one of de Sitter Reissner-Nordstrom black hole. And similar to the de Sitter Reissner-Nordstrom black hole, the charge may enhance the instability. However, the presence of the Gauss-Bonnet term does make difference. When the cosmological constant is small, the term helps to stabilize the black hole such that the black hole could be stable no matter how large the charge is. When the cosmological constant is large, on the contrary the term helps to destabilize the black hole such that the unstable region is enlarged. When the Gauss-Bonnet coefficient is very large, the stability of the black
hole is  determined by a critical cosmological constant which is independent of the charge.


\item At the marginal lines of instabilities there exists  a non-trivial zero-mode static perturbation. This suggests the existence of a non-spherically symmetric static charged de Sitter Gauss-Bonnet black hole solution branch, which can be constructed  by solving the large $D$ effective equations.
\end{enumerate}

On the other hand, if the Gauss-Bonnet term is negative, the stability of the black hole is slightly more complex.  The presence of such a negative term would  make the black holes unstable. It turns out that the Gauss-Bonnet black hole could be unstable no matter the spacetime is asymptotically flat or (A)dS. Especially,  if $\overline{\alpha}<-1/2$, the asymptotically flat and AdS Gauss-Bonnet black holes are always unstable for the whole range of the parameters. But as the parameter $\overline{\alpha}$ turns to zero, the unstable regions shrink to zero for the asymptotically flat and AdS black holes. Furthermore, similar to the case with a positive Gauss-Bonnet term,  there exists  a non-spherically symmetric static black hole solution branch as well.

 The work in this paper can be extended in several directions. For example, it is possible to study Gauss-Bonnet black branes by using the large $D$ expansion method.
 Up to now the Gauss-Bonnet black brane has not been known in a closed form. The large $D$ expansion method may provide a potential tool to construct the analytical solution  and help to study  the phase structure  and the Gregory-Laflamme instability\cite{GL1993,GL1994}. The similar discussions have already been made in the  Einstein gravity \cite{Emparan:2015gva, ST:Nonuniform, EILST:Hydro, RVE:Onbrane, SS:Nonsph}. It might also be interesting to extend the study to the rotating cases, just like the ones  in the Einstein gravity \cite{ST, EST1402,Tanabe:2016opw}. In the Gauss-Bonnet gravity, the construction of the rotating black holes is still an open question. We wish the large $D$ analysis may shed light on this interesting issue. Moreover, it is certainly interesting to address various issues of the Gauss-Bonnet black holes in the framework of membrane paradigm developed in \cite{membrane,chargedmembrane,Dandekar:2016fvw,Dandekar:2016jrp,Bhattacharyya:2016nhn}.


\vspace*{10mm}
\noindent {\large{\bf Acknowledgments}}\\

The work was in part supported by NSFC Grant No.~11275010, No.~11335012 and No.~11325522.


\begin{thebibliography}{}
\bibitem{EST13} R. Emparan, R. Suzuki, and K. Tanabe,{\em  The large $D$ limit of General Relativity,} JHEP \textbf{1306} (2013) 009 [arXiv:1302.6382[hep-th]].
\bibitem{EST14}R.~Emparan, D.~Grumiller and K.~Tanabe,
  {\em Large-D gravity and low-D strings,}
  Phys.\ Rev.\ Lett.\  {\bf 110}, no. 25, 251102 (2013)
  [arXiv:1303.1995 [hep-th]].\\
  R.~Emparan and K.~Tanabe, {\em Universal quasinormal modes of large D black holes,}  Phys.\ Rev.\ D {\bf 89}, no. 6, 064028 (2014)
  [arXiv:1401.1957 [hep-th]].\\
  R.~Emparan, R.~Suzuki and K.~Tanabe,
  {\em Quasinormal modes of (Anti-)de Sitter black holes in the 1/D expansion,}
  JHEP {\bf 1504}, 085 (2015)
  [arXiv:1502.02820 [hep-th]].
\bibitem{EST1406} R. Emparan, R. Suzuki and K. Tanabe, {\em Decoupling and non-decoupling dynamics of large D black holes,} JHEP \textbf{1407}, 113 (2014) [arXiv:1406.1258 [hep-th]].

\bibitem{ESTT} R. Emparan, T. Shiromizu, R. Suzuki, K. Tanabe, and T. Tanaka, {\em Effective theory of Black Holes in the 1/D expansion,} JHEP \textbf{06} (2015) 159, [arXiv:1504.06489 [hep-th]].
\bibitem{ST} R. Suzuki and K. Tanabe, {\em Stationary black holes: Large D analysis,} JHEP \textbf{1509}, 193(2015) [arXiv:1505.01282 [hep-th]].
\bibitem{BDMMS} S. Bhattacharyya, A. De, S. Minwalla, R. Mohan and A. Saha, {\em A membrane paradigm at large D,} JHEP \textbf{1604}, 076 (2016) [arXiv:1504.06613 [hep-th]].
\bibitem{BMMT}S. Bhattacharyya, M. Mandlik, S. Minwalla and S. Thakur, {\em A Charged Membrane Paradigm at Large D,} JHEP \textbf{1604}, 128 (2016) [arXiv:1511.03432 [hep-th]].

\bibitem{Emparan:2015gva}  R.~Emparan, R.~Suzuki and K.~Tanabe,  {\em Evolution and End Point of the Black String Instability: Large D Solution,}
  Phys.\ Rev.\ Lett.\  {\bf 115}, no. 9, 091102 (2015)
   [arXiv:1506.06772 [hep-th]].

\bibitem{Tanabe:2015hda}  K.~Tanabe,  {\em Black rings at large D,}  JHEP {\bf 1602}, 151 (2016)  [arXiv:1510.02200 [hep-th]].
\bibitem{ST:Nonuniform}R. Suzuki and K. Tanabe, {\em Non-uniform black strings and the critical dimension in the
1/D expansion,} JHEP 10 (2015) 107, [arXiv:1506.01890].
\bibitem{EILST:Hydro}R Emparan, K Izumi, R Luna, R Suzuki, K Tanabe, {\em Hydro-elastic
complementarity in black branes at large D,} JHEP \textbf{06} (2016) 117, [arXiv:1602.05752].
\bibitem{RVE:Onbrane} M. Rozali and A. Vincart-Emard, {\em On Brane Instabilities in the Large D Limit,} DOI:
10.1007/JHEP08(2016)166 [arXiv:1607.01747].


\bibitem{Zwiebach}  B.~Zwiebach,  {\em Curvature Squared Terms and String Theories,}  Phys.\ Lett.\ B {\bf 156}, 315 (1985).
\bibitem{Boulware} D. G. Boulware and S. Deser, {\em String-Generated Gravity Models, } Phys. Rev. Lett.\textbf{55} (1985), no. 24 2656.
\bibitem{Wheeler} J. T. Wheeler, {\em Symmetric solutions to the Gauss-Bonnet extended Einstein equations,} Nucl. Phys. \textbf{B268}, 737 (1986).
\bibitem{Wiltshire} D. Wiltshire, {\em Spherically symmetric solutions of Einstein-Maxwell theory with a Gauss-Bonnet term,} Phys. Lett. \textbf{B169}, 36 (1986).
\bibitem{Wiltshire1988} D. Wiltshire, {\em Black holes in string-generated gravity models,} Phys. Rev. \textbf{D38} 2445 (1988).
\bibitem{CFLY} B. Chen, Z. Y. Fan, P. Li and W. Ye,{\em Quasinormal modes of Gauss-Bonnet black holes at large D,} JHEP \textbf{01} 085 (2016) arXiv:1511.08706.

\bibitem{Dotti} G. Dotti and R. J. Gleiser, {\em Linear stability of Einstein-Gauss-Bonnet static spacetimes. Part I: tensor perturbations,} Phys. Rev. \textbf{D72}, 044018 (2005) [arXiv:gr-qc/0503117].
\bibitem{Gleiser} R. J. Gleiser and G. Dotti, {\em Linear stability of Einstein-Gauss-Bonnet static spacetimes. Part II: vector and scalar perturbations,} Phys. Rev. \textbf{D72}, 124002 (2005) [arXiv:gr-qc/0510069].
\bibitem{Konoplya} R. A. Konoplya and A. Zhidenko, {\em (In)stability of D-dimensional black holes in Gauss-Bonnet theory,} Phys. Rev. \textbf{D77}, 104004 (2008) [arXiv:0802.0267 [hep-th]].
\bibitem{Cuyubamba} M. A. Cuyubamba, R. A. Konoplya, A. Zhidenko, {\em Quasinormal modes and a new instability of Einstein-Gauss-Bonnet black holes in the de Sitter world,} Phys. Rev. \textbf{D93}, 104053 (2016) [arXiv:1604.03604 [hep-th]].
\bibitem{Tanabe15} K. Tanabe, {\em Instability of de Sitter Reissner-Nordstrom black hole in the 1/D expansion,} Class. Quant. Grav. \textbf{33} no. 12, 125016 (2016) [arXiv:1511.06059 [hep-th]].
\bibitem{MP} R. C. Myers and M. J. Perry, {\em Black Holes In Higher Dimensional Space-Times,} Annals Phys. \textbf{172}, 304 (1986).
\bibitem{Chen:2017wpf}  B.~Chen, P.~C.~Li and Z.~z.~Wang,   {\em Charged Black Rings at large D,}  arXiv:1702.00886 [hep-th].





\bibitem{Konoplya:2008au}
  R.~A.~Konoplya and A.~Zhidenko,
  {\em Instability of higher dimensional charged black holes in the de-Sitter world,}
  Phys.\ Rev.\ Lett.\  {\bf 103}, 161101 (2009)
  [arXiv:0809.2822 [hep-th]].
  \bibitem{GL1993} R. Gregory and R. Laflamme, {\em Black strings and p-branes are unstable,} Phys. Rev. Lett. \textbf{70} (1993) 2837 [hep-th/9301052].
\bibitem{GL1994}  R. Gregory and R. Laflamme, {\em The Instability of charged black strings and p-branes,}
Nucl. Phys. \textbf{B428} (1994) 399 [hep-th/9404071].
\bibitem{SS:Nonsph}A. Sadhu and V. Suneeta,{\em Non-spherically symmetric black string perturbations in the large D limit,} Phys. Rev. D \textbf{93},124002 (2016)[arXiv:1604.00595].
\bibitem{EST1402} R. Emparan, R. Suzuki, and K. Tanabe, {\em Instability of rotating black holes: large D analysis,} JHEP \textbf{1406} (2014) 106, [arXiv:1402.6215 [hep-th]].


\bibitem{Tanabe:2016opw}  K.~Tanabe,  {\em Charged rotating black holes at large D,}  [arXiv:1605.08854 [hep-th]].

\bibitem{membrane}S. Bhattacharyya, A. De, S. Minwalla, R. Mohan and A. Saha, ¡°A membrane paradigm at large D,¡± JHEP \textbf{1604}, 076 (2016) [arXiv:1504.06613 [hep-th]].
\bibitem{chargedmembrane} S. Bhattacharyya, M. Mandlik, S. Minwalla and S. Thakur, ¡°A Charged Membrane Paradigm at Large D,¡± JHEP \textbf{1604}, 128 (2016) [arXiv:1511.03432 [hep-th]].
\bibitem{Dandekar:2016fvw}
  Y.~Dandekar, A.~De, S.~Mazumdar, S.~Minwalla and A.~Saha,
  ``The large D black hole Membrane Paradigm at first subleading order,''
  JHEP {\bf 1612}, 113 (2016)
  doi:10.1007/JHEP12(2016)113
  [arXiv:1607.06475 [hep-th]].
\bibitem{Dandekar:2016jrp}
  Y.~Dandekar, S.~Mazumdar, S.~Minwalla and A.~Saha,
  ``Unstable `black branes' from scaled membranes at large $D$,''
  JHEP {\bf 1612}, 140 (2016)
  doi:10.1007/JHEP12(2016)140
  [arXiv:1609.02912 [hep-th]].
\bibitem{Bhattacharyya:2016nhn}
  S.~Bhattacharyya, A.~K.~Mandal, M.~Mandlik, U.~Mehta, S.~Minwalla, U.~Sharma and S.~Thakur,
  ``Currents and Radiation from the large $D$ Black Hole Membrane,''
  arXiv:1611.09310 [hep-th].
\end{thebibliography}
\end{document}